\documentclass[aip,reprint
]{revtex4-1}

\usepackage{graphicx}
\usepackage{dcolumn}
\usepackage{bm}
\usepackage[table]{xcolor}
\usepackage[utf8]{inputenc}
\usepackage[T1]{fontenc}
\usepackage{mathptmx}
\usepackage{etoolbox}
\usepackage{lineno}

\usepackage{comment}
\usepackage{amsmath, amssymb} 
\usepackage[ruled,vlined]{algorithm2e}

\usepackage{multirow}
\usepackage{subcaption}
\usepackage{soul}
\makeatletter
\def\@email#1#2{
 \endgroup
 \patchcmd{\titleblock@produce}
  {\frontmatter@RRAPformat}
  {\frontmatter@RRAPformat{\produce@RRAP{*#1\href{mailto:#2}{#2}}}\frontmatter@RRAPformat}
  {}{}
}
\makeatother
\begin{document}


\title[]{Dual guidance: ROM-informed field reconstruction with generative models}

\author{Sajad Salavatidezfouli}
\author{Henrik Karstoft}
\affiliation{Department of Electrical and Computer Engineering, Aarhus University, 8200 Aarhus N, Denmark}
\author{Alexandros Iosifidis}
\affiliation{Faculty of Information Technology and Communication Sciences, Tampere University, Tampere, Finland}
\author{Mahdi Abkar}
\email{abkar@mpe.au.dk} 
\affiliation{Department of Mechanical and Production Engineering, Aarhus University, 8200 Aarhus N, Denmark}

\date{\today}

\begin{abstract}
We present a dual-guided framework for reconstructing unsteady incompressible flow fields using sparse observations. The approach combines optimized sensor placement with a physics-informed guided generative model. Sensor locations are selected using mutual information theory applied to a reduced-order model of the flow, enabling efficient identification of high-information observation points with minimal computational cost. These sensors, once selected, provide targeted observations that guide a denoising diffusion probabilistic model conditioned by physical constraints. Extensive experiments on laminar cylinder wake flows demonstrate that under sparse sensing conditions, the structured sensor layouts fail to capture key flow dynamics, yielding high reconstruction errors. In contrast, our optimized sensor placement strategy achieves accurate reconstructions with errors as low as 0.05, even with a limited number of sensors, confirming the effectiveness of the proposed approach in data-limited regimes. When the number of sensors is higher than a threshold, however, both methods perform comparably. Our dual-guided approach bridges reduced-order model-based sensor position optimization with modern generative modeling, providing accurate, physics-consistent reconstruction from sparse data for scientific machine-learning problems.
\end{abstract}

\maketitle

\section{\label{sec1}Introduction}
The problem of field reconstruction, i.e., recovering a full spatial field from sparse observations, has posed a longstanding challenge across scientific and engineering problems. A wide range of techniques has been developed to tackle this, spanning from classical interpolation methods, e.g., linear, cubic, or kriging interpolation, to advanced reduced-order modeling approaches such as gappy Proper Orthogonal Decomposition (POD), to data assimilation frameworks like iterative Kalman filtering. More recently, Deep Learning (DL) techniques including super-resolution networks, autoencoders, Physics-Informed Neural Networks (PINNs), and generative models have demonstrated improved performance, particularly for complex, non-linear flows \cite{huang2024diffusionpde, lin2020kriging, sekar2022overcoming, fernandez2023generative, hosseini2024flow, shu2023physics}.

Generative models are algorithms that learn the underlying probability distribution of a dataset, allowing them to produce new samples that are statistically consistent with the training data. These models map random noise to realistic data through stochastic processes and can be conditioned on observed fields, allowing for targeted generation that incorporates known information \cite{zhou2023denoising}. Among generative approaches, Generative Adversarial Networks (GANs) have been widely explored for flow field reconstruction. A notable work by Buzzicotti et al. \cite{buzzicotti2021reconstruction} demonstrated the use of GANs to reconstruct 2D slices of 3D rotating turbulence, even in the presence of large gaps and missing features across multiple scales, achieving promising levels of accuracy. However, GANs often suffer from limitations such as mode collapse, unstable training dynamics, and difficulties in capturing fine-grained physical details that can hinder their performance in high-fidelity scientific reconstruction tasks \cite{zhang2018convergence, yousif2024flow}. 

To overcome these limitations, alternative generative frameworks have been explored. Among them, Denoising Diffusion Probabilistic Models (DDPMs) have emerged as a promising candidate. These models gradually transform random noise into realistic data through a sequence of denoising steps governed by a learned reverse diffusion process. Unlike GANs, DDPMs offer stable training and a principled likelihood-based formulation. Recent studies have shown that DDPMs consistently outperform GANs in image synthesis tasks within engineering and scientific domains, offering greater robustness and fidelity in reconstructing complex physical fields \cite{dhariwal2021diffusion,mukhopadhyay2023diffusion, bayat2023study}.

Several studies have employed diffusion models to address both forward problems \cite{jacobsen2025cocogen} and inverse problems \cite{bastek2024physics, dasgupta2025conditional}, with some incorporating physical losses to improve the fidelity of the generated fields \cite{Guo2024EnhancingHR, Bastek2024PhysicsInformedDM}. These approaches are typically built upon either full diffusion or latent diffusion frameworks, where the optimization process involves backpropagation through physics-based loss functions derived from governing Partial Differential Equations (PDEs) as well as data-driven losses based on partial observations \cite{du2024confild, gao2025generative}.

Field reconstruction using diffusion models is primarily achieved through inpainting and guided sampling techniques \cite{song2023pseudoinverse, mardani2023variational, chung2022diffusion}. Inpainting is typically employed for tasks such as super-resolution or the prediction of missing variables, while guided sampling incorporates physical constraints to reconstruct unknown regions of a field based on available observations \cite{hu2024generative}. A key factor influencing the performance of guided diffusion models is the placement of these observations, referred to hereafter as sensor placement. The accuracy of reconstruction is highly dependent on where sensors are located. However, to the best of our knowledge, the optimization or guidance of sensor placement has not yet been systematically explored in the context of diffusion models for field reconstruction.

Sensor placement in a transient field is a complex but critical problem, especially for tasks like flow reconstruction, control, or state estimation. Effective sensor configuration requires accounting not only for spatial distribution but also for temporal dynamics. In particular, one must identify regions that are rich in unique, non-redundant information about the system, avoiding overly correlated measurements and instead targeting locations where critical flow features such as vortices, shear layers, or separation points evolve. 

Several data-driven approaches to sensor placement leverage the observation that fluid flow dynamics can often be approximated in a low-dimensional subspace. Among these, gappy POD \cite{willcox2006unsteady, li2024data} selects sensor locations that best recover the modal coefficients of the flow, while Discrete Empirical Interpolation Methods (DEIM) offer sampling strategies aimed at the efficient reconstruction of nonlinear terms in Reduced-Order Models (ROMs) \cite{drmac2016new}. Building on these ideas, Manohar et al. \cite{manohar2018data} introduced Sparse Sensor Placement Optimization for Reconstruction (SSPOR) method, which extends the Empirical Interpolation Method (EIM) to permit oversampling where the number of sensors exceeds the dimensionality of the reduced model, resulting in improved reconstruction accuracy compared to DEIM. Additionally, various compressive sensing-based techniques have been developed to enable signal recovery from sparse measurements \cite{Rao2024InversePE, Huang2024CompressedSB}.

In this work, we address the challenge of sensor placement for field reconstruction by introducing a novel method based on Mutual Information (MI) theory. While MI provides a principled framework to quantify the information shared between spatial locations, thereby enabling the identification of maximally informative and minimally redundant sensor points, its direct application becomes computationally intractable for high-dimensional, real-world problems such as fluid flows, where the number of candidate locations is large \cite{Zhu2025DistributedPS}. To overcome this, we propose an efficient approach that leverages mutual information between the full-order representation of the flow and a reduced-order solution manifold. This formulation allows us to identify spatial locations that carry the highest mutual information between the full and low-dimensional representations of the system, effectively selecting sensor locations that preserve the essential dynamics of the flow. By doing so, we explicitly guide both sensor placement and the generative reconstruction process, aligning them with flow dynamics and physical laws. This approach scales efficiently with domain size while preserving dominant flow features, as informed by the underlying physics. The rest of the paper is organized as follows. Section \ref{sec2} introduces our proposed methodology, including the MI-based sensor placement strategy and the guided denoising diffusion model for field reconstruction. Section \ref{sec3} details the data generation process used to train and evaluate the model, including the computational setup and design of experiments. Section \ref{sec4} presents and discusses the experimental results, comparing the reconstruction accuracy of structured and optimized sensor layouts. Finally, Section \ref{sec5} concludes the paper and outlines potential directions for future work.
\section{\label{sec2}Methodology}
Our proposed methodology consists of two key components: MI-based sensor placement and a guided denoising diffusion model for field reconstruction. We begin by formulating the sensor location optimization, aiming to identify spatial locations that maximize the mutual information between the full-order physical field and a reduced-order representation. This approach ensures that selected sensors capture the most informative and least redundant aspects of the system's dynamics, while maintaining computational tractability. Once the sensor position is established, we employ a guided DDPM to reconstruct the high-fidelity field from sparse sensor measurements. The generative model is conditioned on the selected sensor values and is guided by a physics-informed prior. The following sections detail the theoretical foundation and implementation of each component.

\subsection{\label{sec2_s_MIT}Mutual Information Theory}
We consider the problem of optimally placing sensors in a high-dimensional physical system, with the goal of maximizing the informativeness of measurements about the underlying field. To this end, we adopt \textit{Mutual Information (MI)} as a model-agnostic, information-theoretic metric. Given two random variables \( \mathbf{u} \) and \( \mathbf{a} \), MI, denoted by \( I(\mathbf{u}; \mathbf{a}) \), quantifies the statistical dependency between them, representing the amount of information that one variable provides about the other.

Formally, the mutual information between \( \mathbf{u} \) and \( \mathbf{a} \), both defined over a probability space, is given by:
\begin{equation}
\label{eq_I1}
\begin{aligned}
I(\mathbf{u}; \mathbf{a}) = \int_{\mathcal{U} \times \mathcal{A}} \log \left( \frac{p(\mathbf{u},\mathbf{a})}{p(\mathbf{u})p(\mathbf{a})} \right) p(\mathbf{u},\mathbf{a}) \, \mathrm{d}\mathbf{u} \, \mathrm{d}\mathbf{a},
\end{aligned}
\end{equation}
where \( p(\mathbf{u},\mathbf{a}) \) is the joint probability density function (PDF) of \( \mathbf{u}\) and \( \mathbf{a} \), and \( p(\mathbf{u}) \), \( p(\mathbf{a}) \) denote their marginal PDFs. Equivalently, this can be expressed as the Kullback–Leibler divergence between the joint distribution and the product of the marginals:
\begin{equation}
\label{eq_I2}
\begin{aligned}
I(\mathbf{u}; \mathbf{a}) = D_{\text{KL}}(p(\mathbf{u},\mathbf{a}) \, \| \, p(\mathbf{u})p(\mathbf{a})).
\end{aligned}
\end{equation}

MI is a non-negative, symmetric measure, satisfying \( I(\mathbf{u};\mathbf{a}) = I(\mathbf{a};\mathbf{u}) \geq 0 \), with equality to zero if and only if \( \mathbf{u} \) and \( \mathbf{a} \) are statistically independent. As such, it provides a natural scalar quantity to assess the informativeness of a sensor measurement \( \mathbf{a} \) with respect to the underlying system state \( \mathbf{u} \). This mutual information framework can be readily applied to quantify the relationship between system states and sensor measurements in high-dimensional physical simulations. In our framework, this metric serves to identify sensor locations that are most informative about reduced-order dynamics. 

In our setting, the full-order solution \( \mathbf{u} \in \mathbb{R}^{m \times n} \) consists of evaluations of the state field at \( m \) spatial locations, across \( n \) realizations drawn from a general parameter space. Each column \( \mathbf{u}(\cdot; \mu_j) \in \mathbb{R}^m \) corresponds to a solution associated with a parameter instance \( \mu_j \in \mathcal{P} \), where \( \mathcal{P} \subset \mathbb{R}^d \) encompasses temporal, physical, or geometric parameters.

To obtain a compact representation of the solution manifold, we perform \textit{Proper Orthogonal Decomposition (POD)} on \( \mathbf{u} \), yielding a reduced-order model of rank \( r \ll m \). The POD decomposition expresses the full field as:
\begin{equation}
\label{eq_POD}
\begin{aligned}
\mathbf{u} \approx \Phi \mathbf{a},
\end{aligned}
\end{equation}
where \( \Phi \in \mathbb{R}^{m \times r} \) is the matrix of dominant orthonormal spatial modes (POD basis), and \( a \in \mathbb{R}^{r \times n} \) contains the associated reduced coordinates, or modal coefficients, for each parameter realization.

The mutual information between the full state \( \mathbf{u} \) and the reduced representation \( \mathbf{a} \) is defined by:
\begin{equation}
\label{eq_I3}
\begin{aligned}
I(\mathbf{u}; \mathbf{a}) = \int_{\mathbb{R}^{m} \times \mathbb{R}^{r}} \log \left( \frac{p(\mathbf{u},\mathbf{a})}{p(\mathbf{u})p(\mathbf{a})} \right) p(\mathbf{u},\mathbf{a}) \, \mathrm{d}\mathbf{u} \, \mathrm{d}\mathbf{a},
\end{aligned}
\end{equation}
and quantifies how well the reduced space \( \text{span}(\Phi) \) captures the statistical structure of the full-order system. In this context, \( I(\mathbf{u};\mathbf{a}) \) measures the information retained when projecting high-fidelity simulations into a low-dimensional basis.

\textit{Entropy} is a fundamental concept in information theory that quantifies uncertainty associated with a random variable. In this setting, MI measures the reduction in uncertainty about the full-order field \( \mathbf{u} \) achieved by observing the reduced coordinates \( \mathbf{a} \), or vice versa. Let \( H(\mathbf{u}) \) denote the \textit{differential entropy} of the random variable \( \mathbf{u} \in \mathbb{R}^m \), which quantifies the average uncertainty (or information content) in the full-order system. Similarly, let \( H(\mathbf{u}|\mathbf{a}) \) denote the \textit{conditional entropy} of \( \mathbf{u} \) given \( \mathbf{a} \), which captures the remaining uncertainty about \( \mathbf{u} \) once the reduced representation \( \mathbf{a} \in \mathbb{R}^r \) is known. Then, the mutual information between \( \mathbf{u} \) and \( \mathbf{a} \) can be equivalently expressed as:
\begin{equation}
\label{eq_H1}
\begin{aligned}
I(\mathbf{u}; \mathbf{a}) = H(\mathbf{u}) - H(\mathbf{u}|\mathbf{a}),
\end{aligned}
\end{equation}
revealing that MI directly quantifies the expected reduction in entropy, i.e., the uncertainty about \( \mathbf{u} \), due to knowledge of \( \mathbf{a} \). This captures how much of the variability in the full-order system is explained by the reduced-order representation.

Assuming that the reduced representation follows a multivariate Gaussian distribution, i.e., \( a \sim \mathcal{N}(0, \Sigma_a) \), and that sensor measurements are linear observations of the full state corrupted by Gaussian noise reflecting inherent measurement uncertainty in the sensing devices, the mutual information admits a closed-form expression that particularly enables optimization. Let \( \mathbf{P} \in \mathbb{R}^{k \times m} \) be the sensor location selection matrix that picks \( k \) rows (sensor locations) from \( \mathbf{u} \). Then the observations are modeled as:
\begin{equation}
\label{eq_y}
\begin{aligned}
\mathbf{y} = \mathbf{P} \mathbf{u} + \eta = \mathbf{P} \Phi \mathbf{a} + \eta,
\end{aligned}
\end{equation}
where \( \eta \sim \mathcal{N}(0, \sigma^2 \mathbf{I}) \) represents additive i.i.d. Gaussian noise. We define the effective observation matrix \( \mathbf{H} = \mathbf{P} \Phi \in \mathbb{R}^{k \times r} \), so that:
\begin{equation}
\label{eq_y2}
\begin{aligned}
\mathbf{y} = \mathbf{H} \mathbf{a} + \eta.
\end{aligned}
\end{equation}

Under this assumption, the MI between the observed state and the latent representation would be \cite{yeung2008differential}:
\begin{equation}
\label{eq_H3}
\begin{aligned}
I(\mathbf{y}; \mathbf{a}) = H(\mathbf{y}) - H(\mathbf{y}|\mathbf{a}),
\end{aligned}
\end{equation}
where both entropies are for multivariate Gaussians and can be written as:
\begin{equation}
\label{eq_H4_2}
\begin{aligned}
H(\mathbf{y}) = \frac{1}{2} \log \det \left( 2\pi e (\mathbf{H} \Sigma_a \mathbf{H}^\top + \sigma^2 \mathbf{I}) \right), \quad 
\end{aligned}
\end{equation}
\begin{equation}
\label{eq_H4}
\begin{aligned}
H(\mathbf{y}|\mathbf{a}) = \frac{1}{2} \log \det \left( 2\pi e \sigma^2 \mathbf{I} \right),
\end{aligned}
\end{equation}
where $e\approx2.718$ stands for the base of the natural logarithm.

Subtracting, the mutual information becomes:
\begin{equation}
\label{eq_H5}
\begin{aligned}
I(\mathbf{y}; \mathbf{a}) = \frac{1}{2} \log \frac{\det(\mathbf{H} \Sigma_a \mathbf{H}^\top + \sigma^2 \mathbf{I})}{\det(\sigma^2 \mathbf{I})}\\
= \frac{1}{2} \log \det \left( \mathbf{I} + \frac{1}{\sigma^2} \mathbf{H} \Sigma_a \mathbf{H}^\top \right).
\end{aligned}
\end{equation}

Finally, using the identity \( \mathbf{H} = \mathbf{P} \Phi \), the mutual information reduces to:
\[
I(\mathbf{y}; \mathbf{a}) = \frac{1}{2} \log \det \left( \mathbf{I} + \frac{1}{\sigma^2} \mathbf{P} \Phi \Sigma_a \Phi^\top \mathbf{P}^\top \right).
\]

This compact expression captures the information gain from measurements in terms of the POD basis \( \Phi \), the prior covariance \( \Sigma_a \), the sensor operator \( \mathbf{P} \), and the noise level \( \sigma^2 \). In essence, it measures how much insight is obtained about the dominant flow structures by sampling the system at specific spatial locations. Moreover, this formulation is particularly well-suited for optimizing sensor placement (encoded by \( \mathbf{P} \)) through greedy algorithms or other discrete optimization techniques.

{\small
\begin{algorithm}[h]
\caption{Greedy Sensor Selection via Mutual Information Maximization}
\KwIn{POD basis matrix $\Phi \in \mathbb{R}^{m \times r}$, covariance $\Sigma_a \in \mathbb{R}^{r \times r}$, noise variance $\sigma^2$, number of sensors $k$}
\KwOut{Optimized sensor index set $\mathcal{S} \subset \{1, \dots, m\}$}

$\mathcal{S} \gets \emptyset$ \tcp*{Initialize the sensor set}

\For{$i = 1$ \KwTo $k$}{
    \ForEach{candidate index $j$ such that $j \notin \mathcal{S}$}{
        \begin{equation*}
        \Delta I_j = \frac{1}{2} \log \det \left( \mathbf{I}_{|\mathcal{S}| + 1} + \frac{1}{\sigma^2} \Phi_{\mathcal{S} \cup \{j\}} \Sigma_a \Phi_{\mathcal{S} \cup \{j\}}^\top \right)
        \end{equation*}  \tcp*{Compute marginal gain}
    }
    
    $j^* \gets \arg\max_{j \notin \mathcal{S}} \Delta I_j$ \tcp*{Index with maximum gain}
    
    $\mathcal{S} \gets \mathcal{S} \cup \{j^*\}$ \tcp*{Update sensor set}
}

\Return{$\mathcal{S}$}
\label{Alg1}
\end{algorithm}
}

\subsubsection{Greedy Algorithm for Sensor Placement}

The optimization problem of selecting \( k \) sensor locations to maximize mutual information is inherently combinatorial. Let \( \mathcal{S} \subseteq \{1, \dots, m\} \) be the index set of the selected sensor locations, where \( m \) is the total number of available spatial points. The goal is to solve:
\begin{equation}
\label{eq_Greedy1}
\begin{aligned}
\max_{|\mathcal{S}| = k} \ I(\mathbf{u}_{\mathcal{S}}; \mathbf{a}) = \frac{1}{2} \log \det \left( \mathbf{I}_{|\mathcal{S}|} + \frac{1}{\sigma^2} \Phi_{\mathcal{S}} \Sigma_a \Phi_{\mathcal{S}}^\top \right),
\end{aligned}
\end{equation}
where \( \Phi_{\mathcal{S}} \in \mathbb{R}^{|\mathcal{S}| \times r} \) denotes the submatrix of POD modes corresponding to the sensor indices in \( \mathcal{S} \), and \( \Sigma_a \in \mathbb{R}^{r \times r} \) is the covariance of the reduced coordinates. The matrix \( \Phi_{\mathcal{S}} \Sigma_a \Phi_{\mathcal{S}}^\top \) reflects the prior covariance of the projected field at the selected locations, and the determinant expression captures the information gain under Gaussian assumptions.

Since the number of possible subsets grows exponentially with \( m \), exhaustive search is computationally infeasible for large systems. Instead, we employ a greedy algorithm that incrementally constructs the sensor set by selecting, at each iteration, the location that offers the largest marginal increase in mutual information.

Let \( \mathcal{S} = \emptyset \) denote the initial empty sensor location set. For each of the \( k \) sensor selections, the algorithm follows the procedure outlined in Algorithm \ref{Alg1}:

1. For each candidate location \( j \notin \mathcal{S} \), compute the \textit{marginal gain in mutual information}:
   \begin{equation}
   \label{eq_GreedyAlg1}
   \begin{aligned}
   \Delta I_j = \frac{1}{2} \log \det \left( \mathbf{I}_{|\mathcal{S}| + 1} + \frac{1}{\sigma^2} \Phi_{\mathcal{S} \cup \{j\}} \Sigma_a \Phi_{\mathcal{S} \cup \{j\}}^\top \right). 
   \end{aligned}
   \end{equation}
2. Select the index \( j^* \) that yields the maximum gain:
   \begin{equation}
   \label{eq_GreedyAlg2}
   \begin{aligned}
   j^* = \arg\max_{j \notin \mathcal{S}} \Delta I_j.
   \end{aligned}
   \end{equation}
3. Update the sensor set:
   \begin{equation}
   \label{eq_GreedyAlg3}
   \begin{aligned}
   \mathcal{S} \leftarrow \mathcal{S} \cup \{j^*\}.
   \end{aligned}
   \end{equation}

After \( k \) iterations, the set \( \mathcal{S} \) contains the selected sensor locations that greedily maximize the mutual information with the reduced representation \( \mathbf{a} \).

Having established an information theory-grounded method for sensor location selection, we now turn to the second pillar of our framework: reconstructing the full field using a physics-aware guided generative model.

\subsection{\label{sec2_s_GDDPM}Guided Diffusion}
Diffusion models generate data by reversing a noise injection process. A deterministic formulation models this as an Ordinary Differential Equation (ODE)-driven trajectory from noise to data, parameterized by a variance schedule $\sigma(t)$ over time $t \in [0, T]$, as detailed in Ref. [\citenum{song2020score}]. The evolution of a sample $\mathbf{x}$ in this denoising process is governed by:
\begin{equation}
d\mathbf{x} = -\dot{\sigma}(t) \sigma(t) \nabla_{\mathbf{x}} \log p(\mathbf{x}; \sigma(t)) dt,
\label{eq:diffusion_ode}
\end{equation}
where $\nabla_{\mathbf{x}} \log p(\mathbf{x}; \sigma(t))$ is the score function, i.e., the gradient of the log-density at noise level $\sigma(t)$, and $\dot{\sigma}(t) $ denotes the time derivative of the noise scale $\sigma(t) $. To estimate this score function, one typically trains a denoiser $D(\mathbf{x}; \sigma)$ such that:
\begin{equation}
\nabla_{\mathbf{x}} \log p(\mathbf{x}; \sigma(t)) = \frac{D(\mathbf{x}; \sigma(t)) - \mathbf{x}}{\sigma(t)^2}.
\label{eq:score_denoiser}
\end{equation}

To steer the generative process toward satisfying the constraints, guided diffusion methods incorporate additional gradients that bias the sampling toward a conditional distribution. In this context, Diffusion Posterior Sampling (DPS)\cite{chung2022diffusion} enables solving inverse problems by conditioning on partial or noisy observations $\mathbf{y}$ derived from the target $\mathbf{x}$.

Incorporating the measurement information into the denoising dynamics modifies Eq.~\eqref{eq:diffusion_ode} as follows:
\begin{equation}
d\mathbf{x} = -\dot{\sigma}(t)\sigma(t) \left( \nabla_{\mathbf{x}} \log p(\mathbf{x}; \sigma(t)) + \nabla_{\mathbf{x}} \log p(\mathbf{y}|\mathbf{x}; \sigma(t)) \right) dt,
\label{eq:guided_diffusion}
\end{equation}
where the second term is the gradient of the likelihood under a measurement model. Assuming a sparse measurement operator, $\mathcal{M}$, the guidance term becomes:
\begin{align}
&\nabla_{\mathbf{x}_i} \log p(\mathbf{y}|\mathbf{x}_i; \sigma(t_i)) \approx \nabla_{{\mathbf{x}}_i} \log p(\mathbf{y}|\hat{\mathbf{x}}^{i}_{N}; \sigma(t_i)) \nonumber\\
&\;\;\;\;\;\;\;\;\;\;\;\;\;\;\;\;\;\;\;\;\;\;\;\approx -\zeta_\text{obs} \nabla_{\mathbf{x}_i} \left || \mathbf{y}_\text{obs} - \mathcal{M}(\hat{\mathbf{x}}^i_N(\mathbf{x}_i; \sigma(t_i))) |\right|_2^2,
\label{eq:gradient_log_likelihood}
\end{align}
where $\hat{\mathbf{x}}^i_N := D(\mathbf{x}_i; \sigma(t_i))$ is the final denoised estimate at each denoising step $i$. Applying Bayes' rule, this leads to an approximation of the posterior score function:
\begin{equation}
\nabla_{\mathbf{x}_i} \log p(\mathbf{x}_i|\mathbf{y}_\text{obs})
\approx s(\mathbf{x}_{i}) - \zeta_\text{obs} \nabla_{\mathbf{x}_i} \left|| \mathbf{y}_\text{obs} - \mathcal{M}(\hat{\mathbf{x}}^i_N) |\right|_2^2,
\label{eq:guided_score}
\end{equation}
where $s(\mathbf{x}_i)$ is the prior score and $\zeta_\text{obs}$ controls the strength of measurement guidance.

To further guide the denoising dynamics toward physically consistent solutions, we incorporate PDE-based guidance into the sampling process. Assuming that the solution $\hat{\mathbf{x}}^i_N := D(\mathbf{x}_i; \sigma(t_i))$ should satisfy a known PDE of the form $f(\hat{\mathbf{x}}^i_N) = 0$, we introduce an additional guidance term derived from the corresponding PDE residual loss. This modifies the posterior score approximation as:
\begin{align}  
&\nabla_{\mathbf{x}_i} \log p(\mathbf{x}_i|\mathbf{y}_\text{obs}, f) \approx \nonumber\\
& \;\; s(\mathbf{x}_{i}) -\zeta_{\text{obs}} \nabla_{\mathbf{x}_i} \left|| \mathbf{y}_\text{obs} - \mathcal{M}(\hat{\mathbf{x}}^i_N)  |\right| _2^2 - \zeta_{\text{pde}} \nabla_{\mathbf{x}_i} \left|| f(\hat{\mathbf{x}}^i_N) |\right|_2^2,
\label{eq:guided_score_pde}
\end{align}
where $\zeta_{\text{pde}}$ controls the influence of the PDE constraint. During inference, the gradients of this residual are backpropagated through the denoising model, enforcing consistency with the underlying physical law represented by the PDE. The sampling process algorithm is presented in Algorithm \ref{Alg2}.

{\tiny
\begin{algorithm}
\caption{Guided Diffusion Sampling Algorithm}
\KwIn{Deterministic Sampler $D_\theta(x; \sigma)$, $\sigma(t_{i \in \{0, \ldots, N\}})$, Observation $\mathbf{y}_\text{obs}$, PDE Function $f$, Guide Weights $\zeta_{\text{obs}}, \zeta_{\text{pde}}$}
\KwOut{Denoised data $x_N$}

Sample $x_0 \sim \mathcal{N}(0, \sigma(t_0)^2\mathbf{I})$ \tcp*{Generate initial sampling noise}

\For{$i = 0$ \KwTo $N - 1$}{
    $\hat{\mathbf{x}}_N^i \gets D_\theta(x_i; \sigma(t_i))$ \tcp*{Estimate denoised data at step $t_i$}
    
    $d_i \gets \frac{x_i - \hat{\mathbf{x}}_N^i}{\sigma(t_i)}$ \tcp*{Evaluate $dx/d\sigma(t)$ at step $t_i$}
    
    $x_{i+1} \gets x_i + \left( \sigma(t_{i+1}) - \sigma(t_i) \right) d_i$ \tcp*{Euler step}
    
    \If{$\sigma(t_{i+1}) \neq 0$}{
        $\hat{\mathbf{x}}_N^i \gets D_\theta(x_{i+1}; \sigma(t_{i+1}))$ \tcp*{2nd order correction}
        
        $d_i' \gets \frac{x_{i+1} - \hat{\mathbf{x}}_N^i}{\sigma(t_{i+1})}$ \tcp*{Evaluate $dx/d\sigma(t)$ at $t_{i+1}$}
        
        $x_{i+1} \gets x_i + \left( \sigma(t_{i+1}) - \sigma(t_i) \right) \cdot \left( \frac{1}{2}d_i + \frac{1}{2}d_i' \right)$ \tcp*{Trapezoidal rule}
    }
    
    $\mathcal{L}_{\text{obs}} \gets  \| \mathbf{y}_\text{obs} - \mathcal{M}(\hat{\mathbf{x}}^i_N) \|_2^2$ \tcp*{Observation loss}
    
    $\mathcal{L}_{\text{pde}} \gets  \| f(\hat{\mathbf{x}}_N^i) \|_2^2$ \tcp*{PDE loss}
    
    $x_{i+1} \gets x_{i+1} - \zeta_{\text{obs}} \nabla_{x_i} \mathcal{L}_{\text{obs}} - \zeta_{\text{pde}} \nabla_{x_i} \mathcal{L}_{\text{pde}}$\;
}

\Return{$x_N$} \tcp*{Return the denoised data}
\label{Alg2}
\end{algorithm}
}

\section{\label{sec3}Data Generation for DDPM Training}
Our approach focuses on solving transient PDEs using a data-driven framework that incorporates guided diffusion processes. Specifically, we focused on the 2D incompressible Navier-Stokes equations as a test case, which are expressed as:
\begin{align}
&\partial_{\tau} \mathbf{v}(\xi, \tau) + \mathbf{v}(\xi, \tau) \cdot \nabla \mathbf{v}(\xi, \tau) \notag + \frac{1}{\rho} \nabla \text{P}\\
&\;\;\;\;\;\;\;\;\;\;\;\;\;\;\;\;\;\;\;\;\;\;\;\;\;\;\;=   \nu \nabla^2 \mathbf{v}(\xi, \tau), \quad \xi \in \Omega, \; \tau \in (0, T],
\label{eq:PDE1}
\end{align}
\begin{align}
&\nabla \cdot \mathbf{v}(\xi, \tau) = 0, \quad \xi \in \Omega, \; \tau \in (0, T],
\label{eq:PDE2}
\end{align}
where, $\tau$ denotes the temporal coordinate, $\Omega$ is the spatial domain, and $\xi \in \Omega$ represents spatial coordinates. 
Guided diffusion is employed to learn the velocity field $\mathbf{v}(\xi, \tau)$ at spatial location $\xi$ and time $\tau$, conditioned on sparse spatial observations and partial knowledge of the governing PDE. While the full Navier-Stokes system comprises both the momentum and continuity equations, enforcing the complete set of constraints would require explicit modeling of the pressure field $\text{P}$, which is non-trivial and introduces additional complexity. To simplify the sampling process, we enforce only the incompressibility condition (Eq. \ref{eq:PDE2}) during sampling. This is incorporated into the loss function to guide backpropagation, encouraging the learned velocity fields to remain divergence-free.

\begin{figure}
    \centering
    \includegraphics[width=0.9\linewidth]{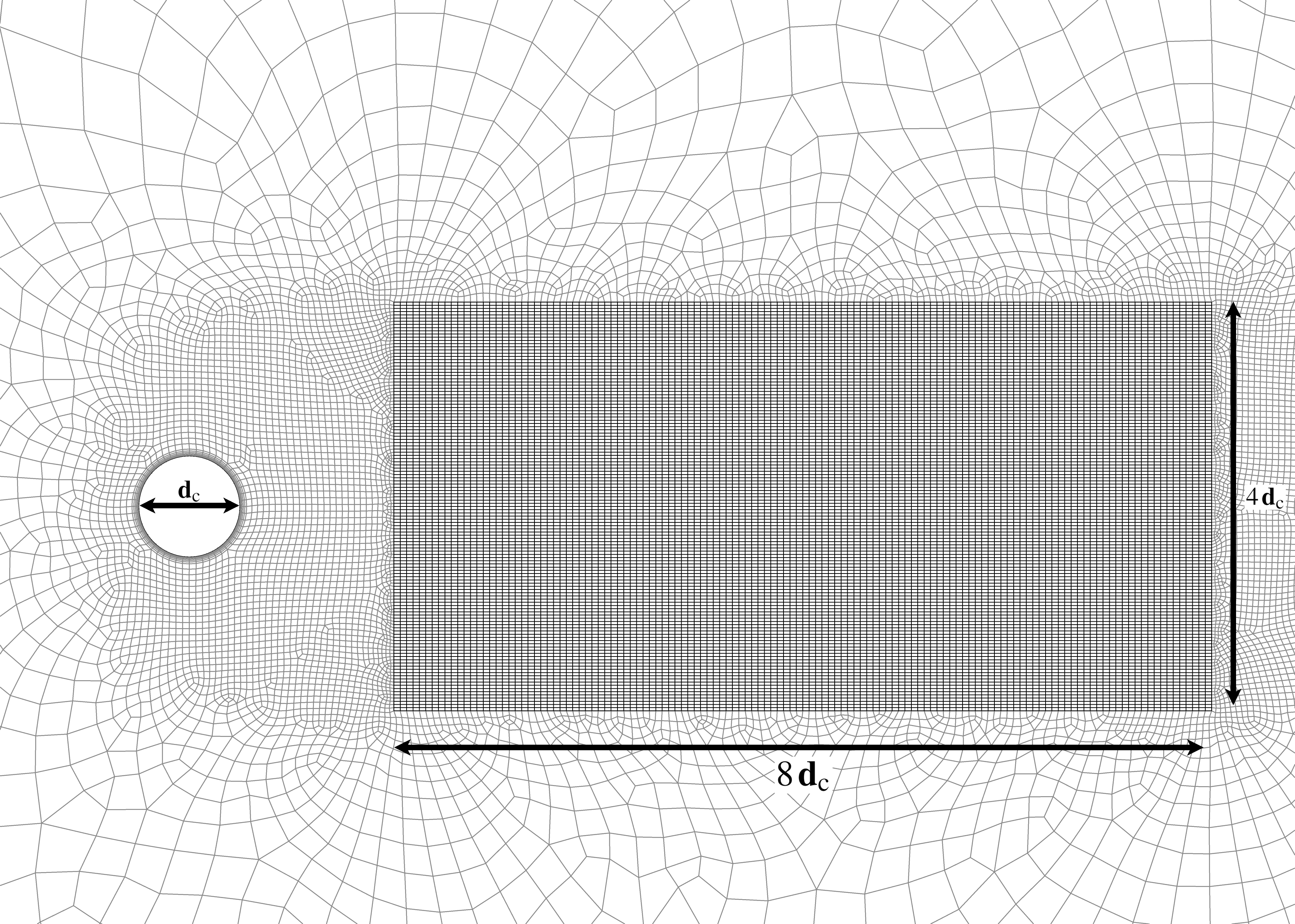}
    \caption{Computational domain surrounding and within the wake region of the circular cylinder. The wake sampling box is shown with dimensions $4\,\mathbf{d_\text{c}} \times 8\,\mathbf{d_\text{c}}$ in this illustration, which vary depending on the case studied. A uniform, structured 128×128 mesh grid is used for all cases within the wake sampling box.}
    \label{fig:SpatialSampling}
\end{figure}

\begin{table*}
\centering
\caption{Simulation properties}
\label{tab:CFDProp}
\small
\begin{tabular}{|c|c|c|c|c|c|}
\hline
$\begin{tabular}[c]{@{}c@{}}Cylinder Diameter\\ $\mathbf{d_\text{c}}$ (m) \end{tabular}$ & Density (kg/m$^3$) & Initial Condition & Solver & \begin{tabular}[c]{@{}c@{}}Computational\\Domain\end{tabular} & \begin{tabular}[c]{@{}c@{}}Wake Grid\\Size\end{tabular} \\
\hline
0.05 & 1.225 & $\mathbf{v}_\text{x}, \mathbf{v}_\text{y}, \text{P} = 0$ & \begin{tabular}[c]{@{}c@{}}Laminar, Incompressible,\\2nd Order Discretization\end{tabular} & $40\mathbf{d_\text{c}} \times 20\mathbf{d_\text{c}}$ & $128 \times 128$ \\
\hline
\end{tabular}
\end{table*}

\begin{table*} \centering 
\caption{Design of Experiment Parameter Space} 
\begin{tabular}{|l|c|c|c|c|} 
\hline 
\rowcolor[gray]{0.9} 
\textbf{Parameter} & \textbf{Min} & \textbf{Max} & \textbf{Values/Increments} & \textbf{Number of Combinations} \\ 
\hline 
Reynolds Number & 50 & 200 & Increment of 5 & 31 \\ 
\hline 
Inlet Velocity (m/s) & 0.01 & 1.00 & 0.01, 0.1, 0.2, 0.3, 0.4, 0.5, 0.6, 0.7, 0.8, 0.9, 1.0 & 11 \\ 
\hline 
Wake Sampling Size & \multicolumn{3}{l|}{ ($3\mathbf{d_\text{c}} \times 6\mathbf{d_\text{c}}$), ($3\mathbf{d_\text{c}} \times 7.5\mathbf{d_\text{c}}$), ($3\mathbf{d\text{c}} \times 9\mathbf{d_\text{c}}$),} & 6 \\ 
& \multicolumn{3}{l|}{ ($4\mathbf{d_\text{c}} \times 6\mathbf{d_\text{c}}$), ($4\mathbf{d_\text{c}} \times 8\mathbf{d_\text{c}}$), ($4\mathbf{d_\text{c}} \times 10\mathbf{d_\text{c}}$)} & \\ 
\hline 
Sampling Frequency (Hz) & \multicolumn{3}{c|}{$100\times\mathcal{F}$ = 100 samples per run} & - \\ 
\hline 
\end{tabular} 
\label{tab:doe_parameters} 
\end{table*}

\subsection{\label{sec3_s_Setup}Flow Configuration and Simulation Setup}
To train the generative diffusion model for field reconstruction, we construct a dataset based on numerical simulations of laminar vortex shedding behind a circular cylinder. The resulting data captures unsteady, spatially coherent structures characteristic of fluid dynamics in the laminar Reynolds number range.

The two-dimensional incompressible laminar flow around a circular cylinder has been examined for Reynolds numbers in the range $50 \leq \text{Re} \leq 200$, corresponding to the periodic vortex shedding regime \cite{zdravkovich1996different}. For all cases, a structured 128×128 mesh grid is employed in the wake region to ensure consistent and accurate data sampling to match the expected input format of the DDPM model, as illustrated in Fig. \ref{fig:SpatialSampling}. To resolve the elongated wake structures, the physical dimensions of the sampling box are chosen to be rectangular, with a larger extent in the streamwise direction than the spanwise, as presented in Table \ref{tab:CFDProp}. This allows higher resolution in regions of coherent vortex dynamics.

Table \ref{tab:CFDProp} presents the main properties of the computational setup. The flow is initialized with zero velocity and pressure fields, and a second-order upwind scheme is used for discretization. The simulation is run until a stable vortex shedding pattern develops.

\subsection{Design of Experiment and Parameter Space}
The DDPM model requires a large number of diverse samples to generalize across variations in physical conditions. To generate $\mathcal{O}(10^4)$ training snapshots, we design a parametric study varying key flow and geometric features:

\begin{itemize}
    \item \textbf{Reynolds Number}: varied from 50 to 200 in steps of 5, yielding 31 unique values. For each Reynolds number, with a fixed cylinder diameter and fluid density, we had the flexibility to adjust the (velocity, viscosity) pair according to the Reynolds equation $\text{Re} = \frac{\rho \mathbf{v}_\text{in} \text{d}_\text{c}}{\mu}$. This approach allowed us to maintain the same Reynolds number while exploring different flow characteristics.
    
    \item \textbf{Inlet Velocity}: varied from $0.01\,\text{m/s}$ to $1.0\,\text{m/s}$ across 11 distinct values. By tuning velocity alongside viscosity while maintaining the same Reynolds number, we captured different vortex morphologies and wake structures. This would significantly enrich our snapshot database despite the vortex shedding Strouhal number remaining constant for a given Reynolds number.
    
    \item \textbf{Temporal Sampling Strategy}: The temporal sampling frequency is based on the vortex shedding period, which varies with Reynolds number. Using empirical fits \cite{roshko1954development, williamson1998series}, the shedding frequency is estimated by the Strouhal number:
    \begin{equation}
    \text{St} = 0.21\left(1 - \frac{21}{\text{Re}}\right), \quad \mathcal{F} = \text{St} \cdot \frac{\mathbf{v}_\text{in}}{\text{d}_\text{c}},
    \label{eq:St_vs_Re}
    \end{equation}
    where $\mathcal{F}$ stands for the vortex shedding frequency.
    
    To ensure sufficient temporal resolution, we extract $100$ snapshots per shedding cycle. The simulation time step is conservatively chosen as $\Delta t_{\text{sim}} = 1/(\mathcal{F}\times200)$, and snapshots are sampled every $2\Delta t_{\text{sim}}$.
  
    \item \textbf{Sampling Box Geometry}: 6 different wake sampling boxes were utilized to provide different spatial perspectives of the flow field, as listed in Table\ \ref{tab:doe_parameters}.
\end{itemize}

This combination of physical and numerical parameters ensures high diversity in the training data. From the parameter space defined above, we generated 300 different simulation cases by randomly selecting combinations of Reynolds numbers, inlet velocities, and domain sizes. Each case was sampled at 100 temporal snapshots per shedding cycle, yielding a comprehensive dataset of 30,000 snapshots for the DDPM training. Table\ \ref{tab:doe_parameters} summarizes the DoE parameters, which govern the sampling strategy for generating synthetic but physically consistent unsteady fields.

\begin{figure}
    \centering
    \includegraphics[width=1\linewidth]{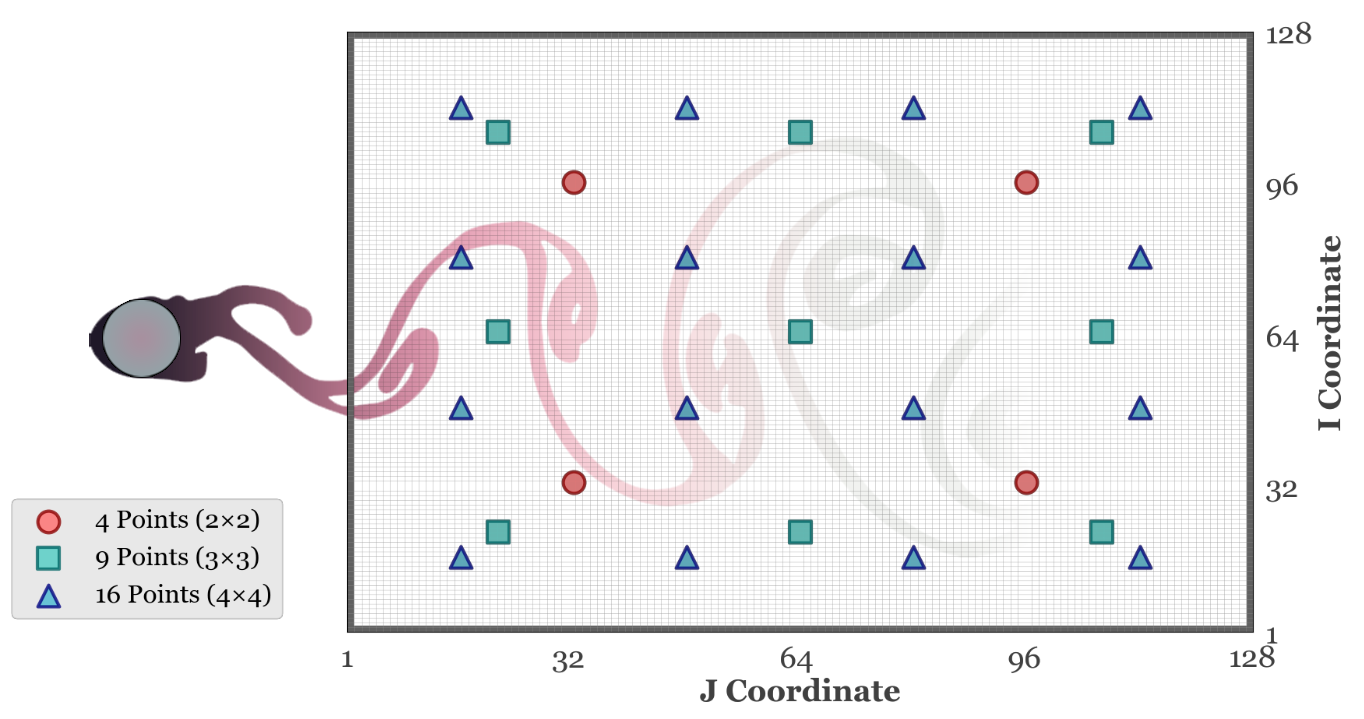}
    \caption{Structured sensor placements in the wake region behind the circular cylinder. The figure shows representative layouts for 4 (2×2), 9 (3×3), and 16 (4×4) sensors, uniformly distributed across the domain.}
    \label{fig:sensor_points_struct}
\end{figure}

\begin{figure}
    \centering
    \includegraphics[width=1\linewidth]{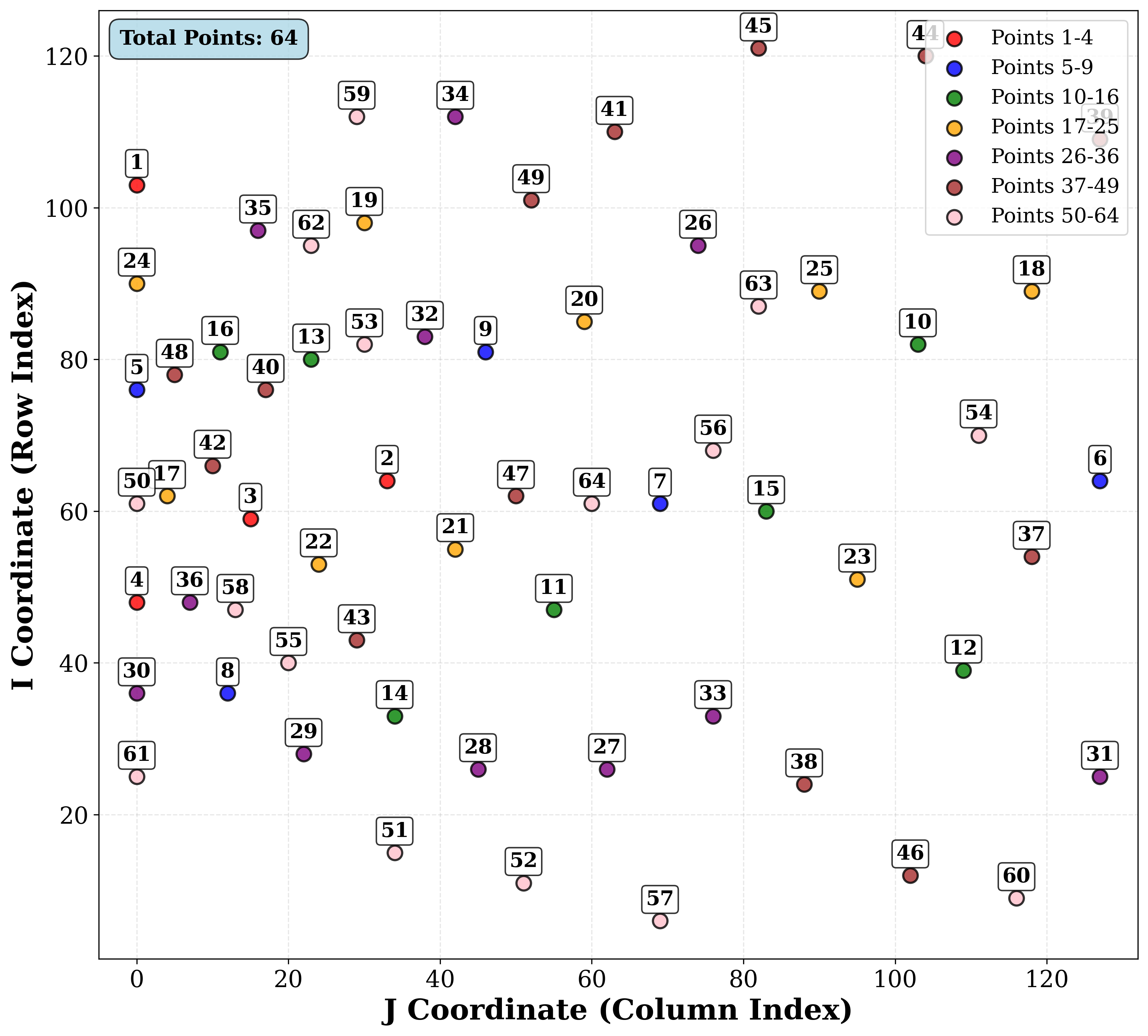}
    \caption{Optimized sensor locations in the wake region of the cylinder, selected using a greedy mutual information algorithm on the reduced-order solution manifold.}
    \label{fig:mutualinfo_points}
\end{figure}

\begin{figure*}
    \centering
    \includegraphics[width=1\linewidth]{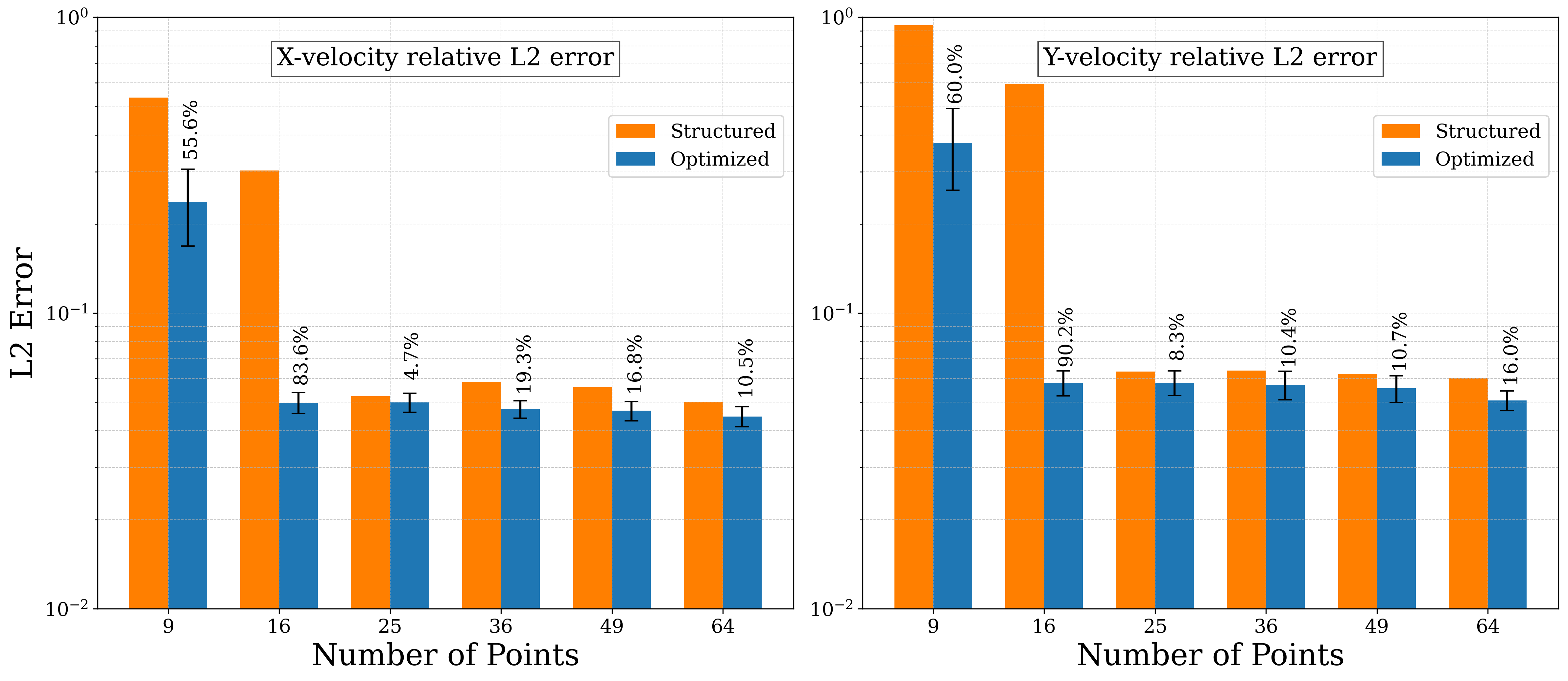}
    \caption{Comparison of total average L2 errors for $\mathbf{v}_\text{x}$ and $\mathbf{v}_\text{y}$ across all test cases and timesteps for structured and optimized sensor placements. Bars represent mean errors with standard deviation shown in black lines, while percentages above the optimized bars denote the relative improvement over structured placement. }
    \label{fig:barimprovement}
\end{figure*}

\section{\label{sec4}Results and Discussion}

As previously mentioned in the data generation section, a total of 300 different simulation cases were performed, each spanning 100 timesteps. For training the guided-DDPM, 250 simulations were utilized, with the remaining 50 reserved for testing. Given the considerable computational cost associated with DDPM sampling (each snapshot requires over five minutes on an NVIDIA A6000 GPU), only four test cases were selected for sampling. Sampling was conducted at four distinct timesteps: 0, 25, 50, and 75, yielding a total of 16 sampling scenarios for each sensor placement strategy to evaluate reconstruction accuracy across different cases, timesteps, and observation points.

To establish a baseline, we first consider structured sensor layouts that uniformly span the wake region. Figure~\ref{fig:sensor_points_struct} illustrates only three sensor resolutions: 2×2 grid (4 sensors), 3×3 grid (9 sensors), and 4×4 grid (16 sensors). Each configuration is represented using different marker shapes and colors for clarity. These sensor locations were selected to provide broad spatial coverage of the domain while maintaining simplicity and interpretability. Although these placements are not tailored to the specific dynamics of the flow, they serve as a valuable reference for evaluating the effectiveness of our advanced placement strategy.

\begin{figure*}
    \centering
    \begin{subfigure}[t]{0.46\textwidth}
        \centering
        \includegraphics[width=\linewidth,trim=0 0 0 0,clip]{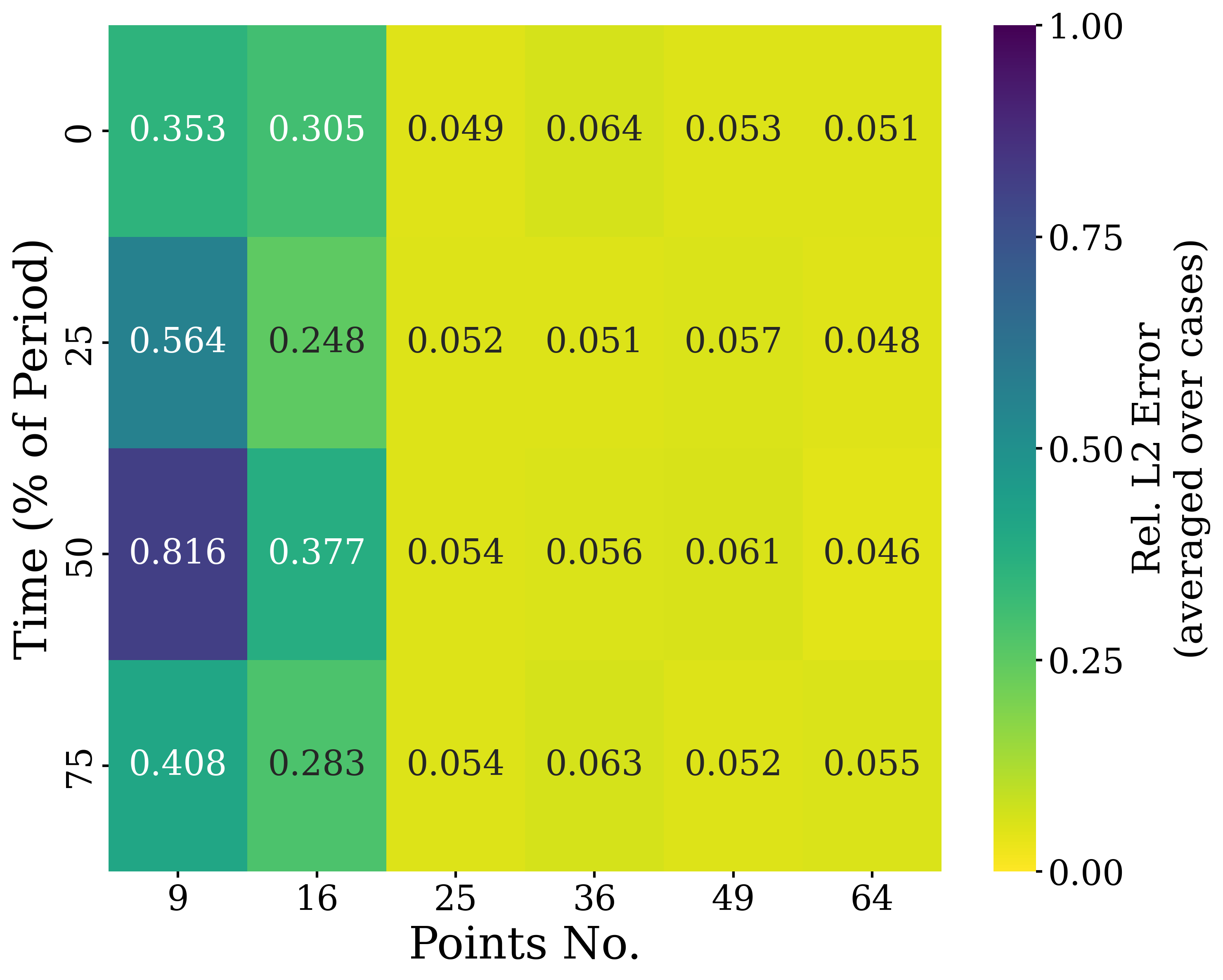}
        \subcaption{$\mathbf{v}_\text{x}$ L2 error with structured sensor placement}
        \label{fig:vx_structured_RUNavg}
    \end{subfigure}
    \hfill
    \begin{subfigure}[t]{0.46\textwidth}
        \centering
        \includegraphics[width=\linewidth,trim=0 0 0 0,clip]{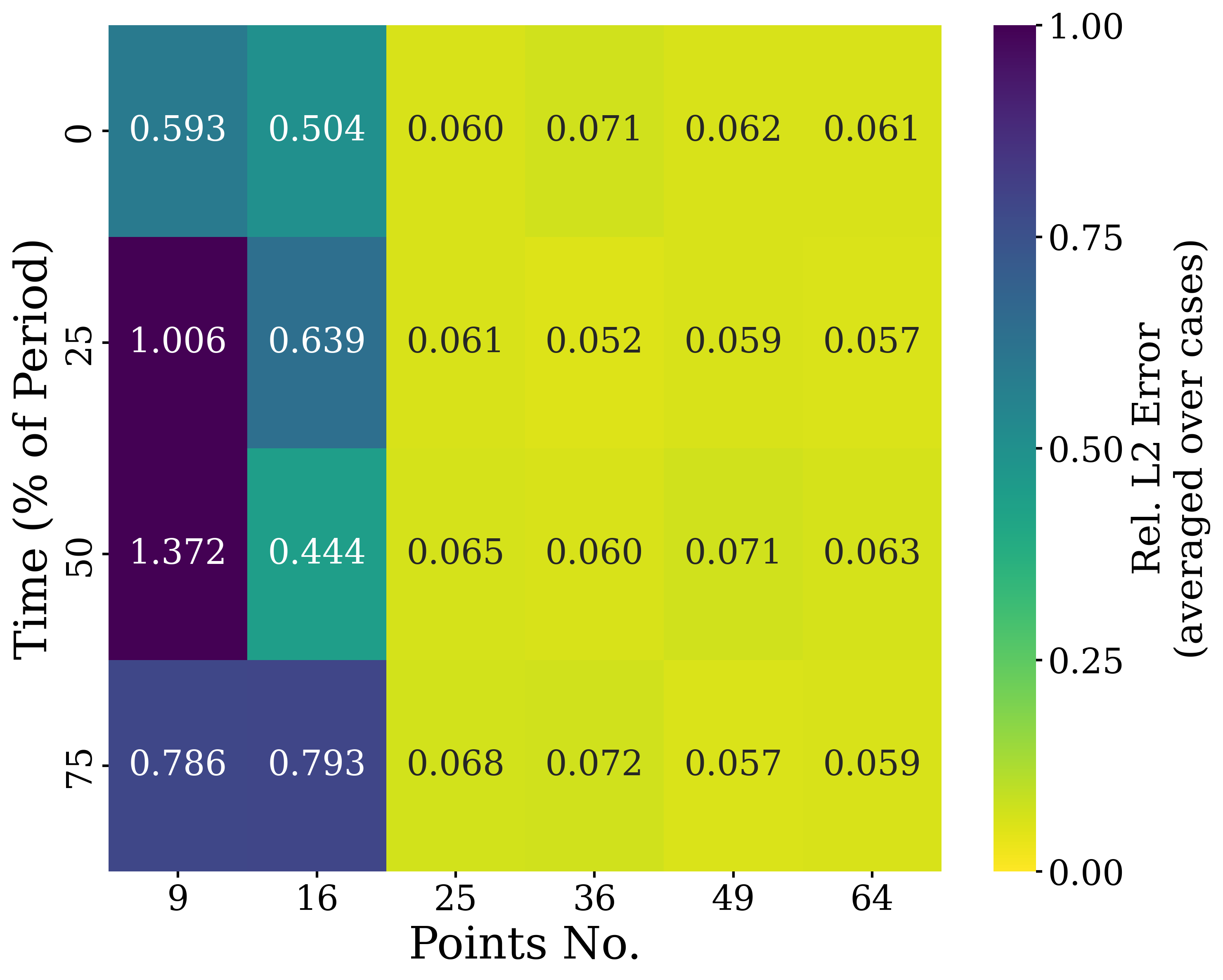}
        \subcaption{$\mathbf{v}_\text{y}$ L2 error with structured sensor placement}
        \label{fig:vy_structured_RUNavg}
    \end{subfigure}
    
    \vspace{0.5cm}
    
    \begin{subfigure}[t]{0.46\textwidth}
        \centering
        \includegraphics[width=\linewidth,trim=0 0 0 0,clip]{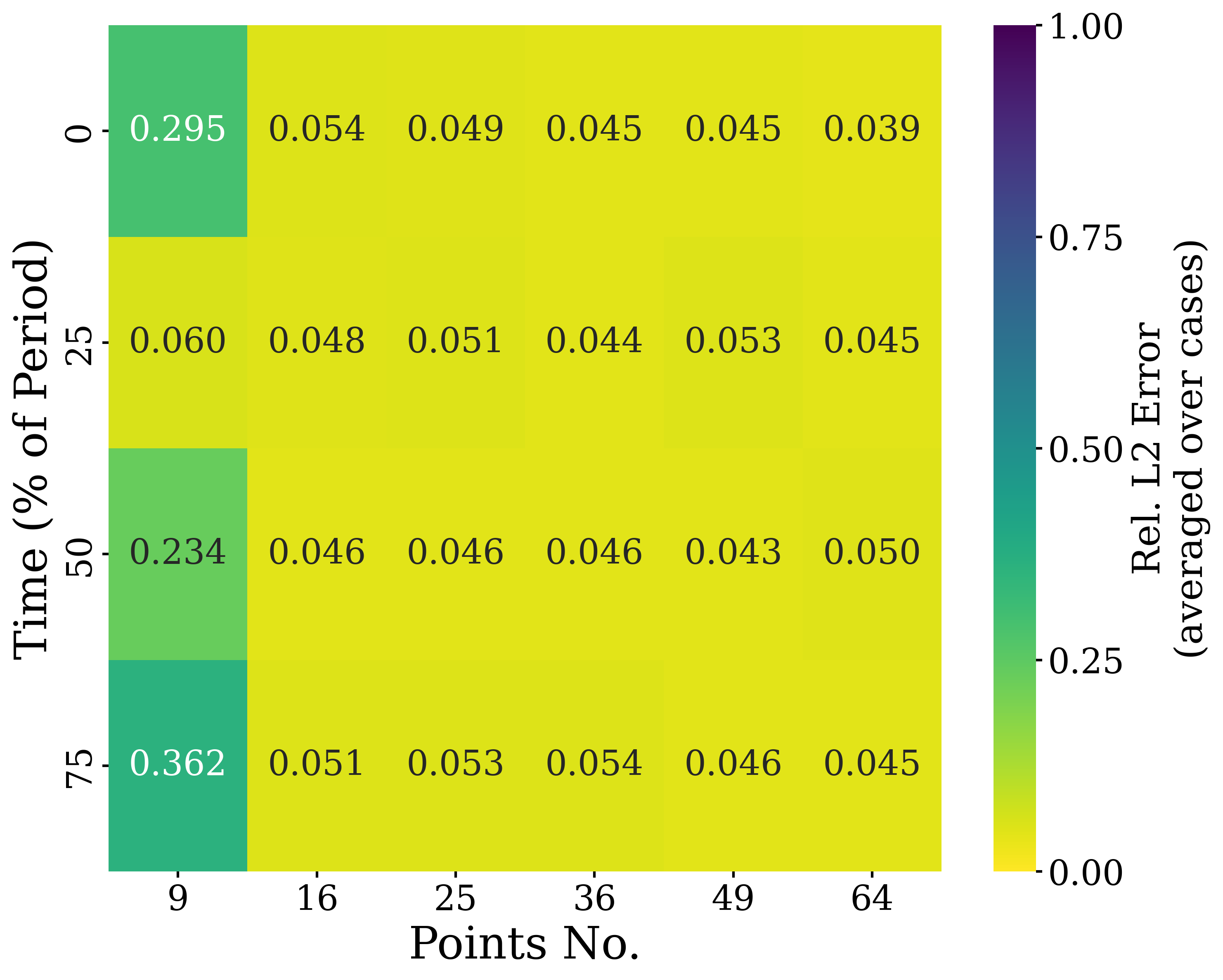}
        \subcaption{$\mathbf{v}_\text{x}$ L2 error with optimized sensor placement}
        \label{fig:vx_optimized_RUNavg}
    \end{subfigure}
    \hfill
    \begin{subfigure}[t]{0.46\textwidth}
        \centering
        \includegraphics[width=\linewidth,trim=0 0 0 0,clip]{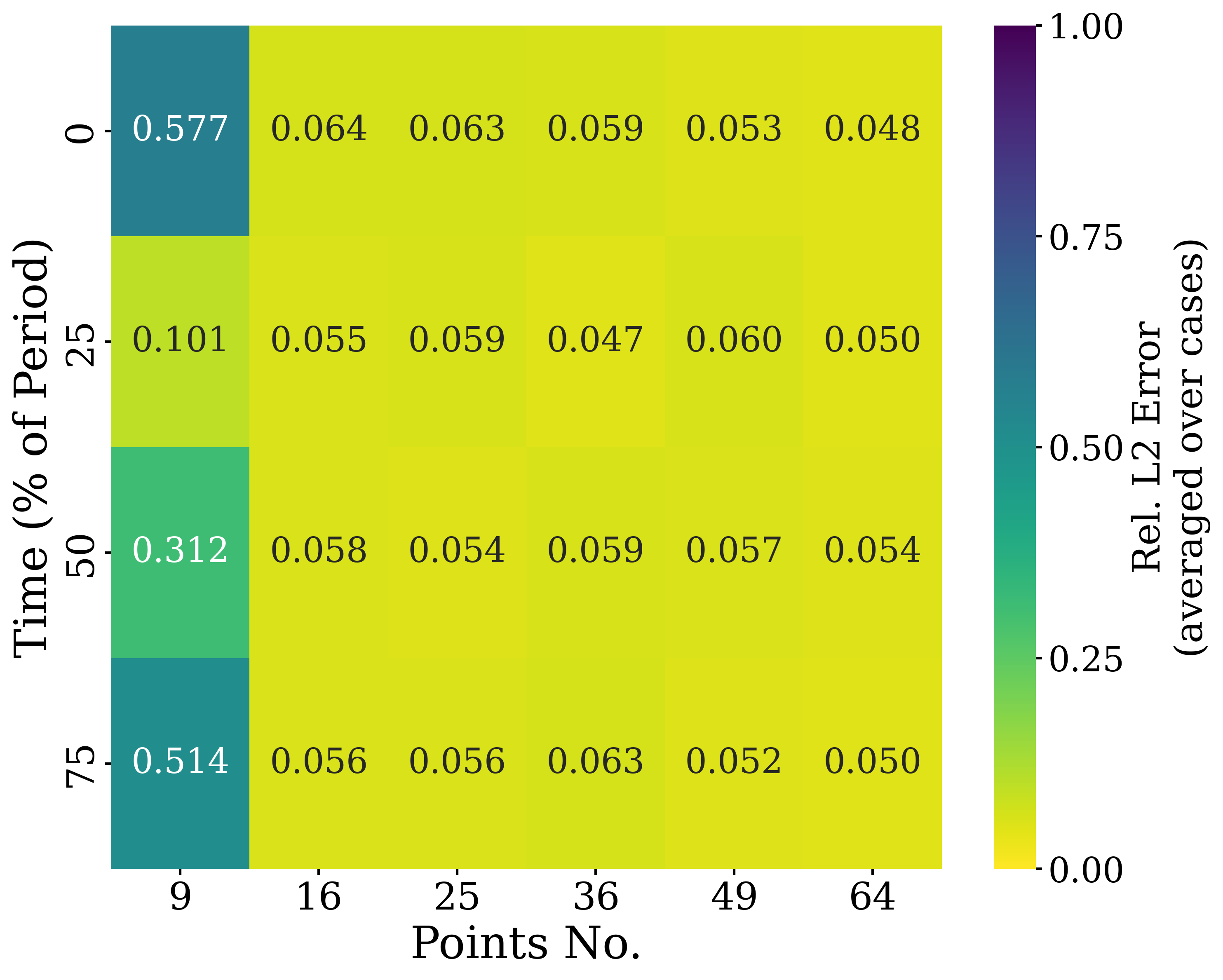}
        \subcaption{$\mathbf{v}_\text{y}$ L2 error with optimized sensor placement}
        \label{fig:vy_optimized_RUNavg}
    \end{subfigure}
    
    \vspace{0.5cm}
    
    \begin{subfigure}[t]{0.46\textwidth}
        \centering
        \includegraphics[width=\linewidth,trim=0 0 0 0,clip]{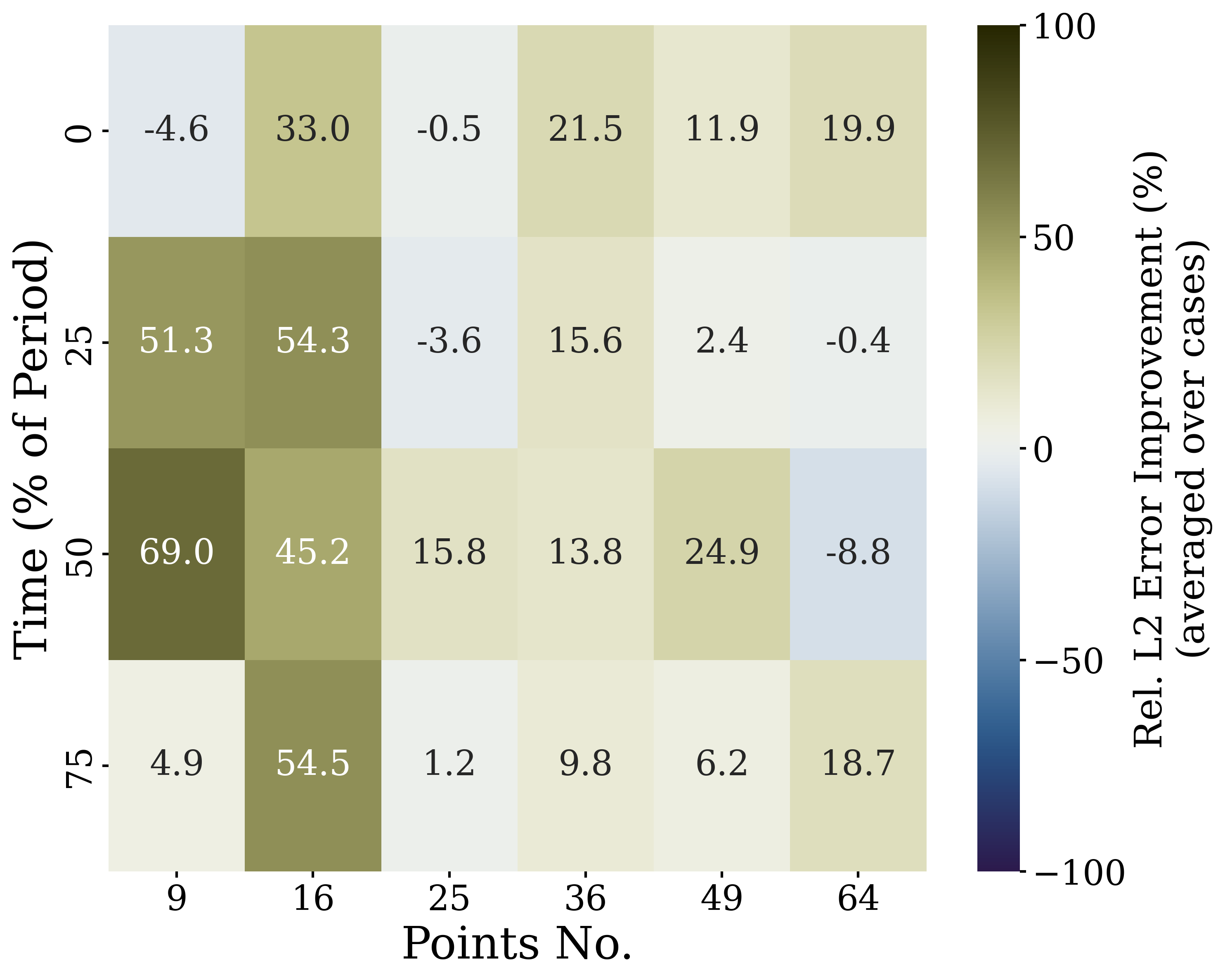}
        \subcaption{L2 error improvement for $\mathbf{v}_\text{x}$ component}
        \label{fig:vx_improvement_RUNavg}
    \end{subfigure}
    \hfill
    \begin{subfigure}[t]{0.46\textwidth}
        \centering
        \includegraphics[width=\linewidth,trim=0 0 0 0,clip]{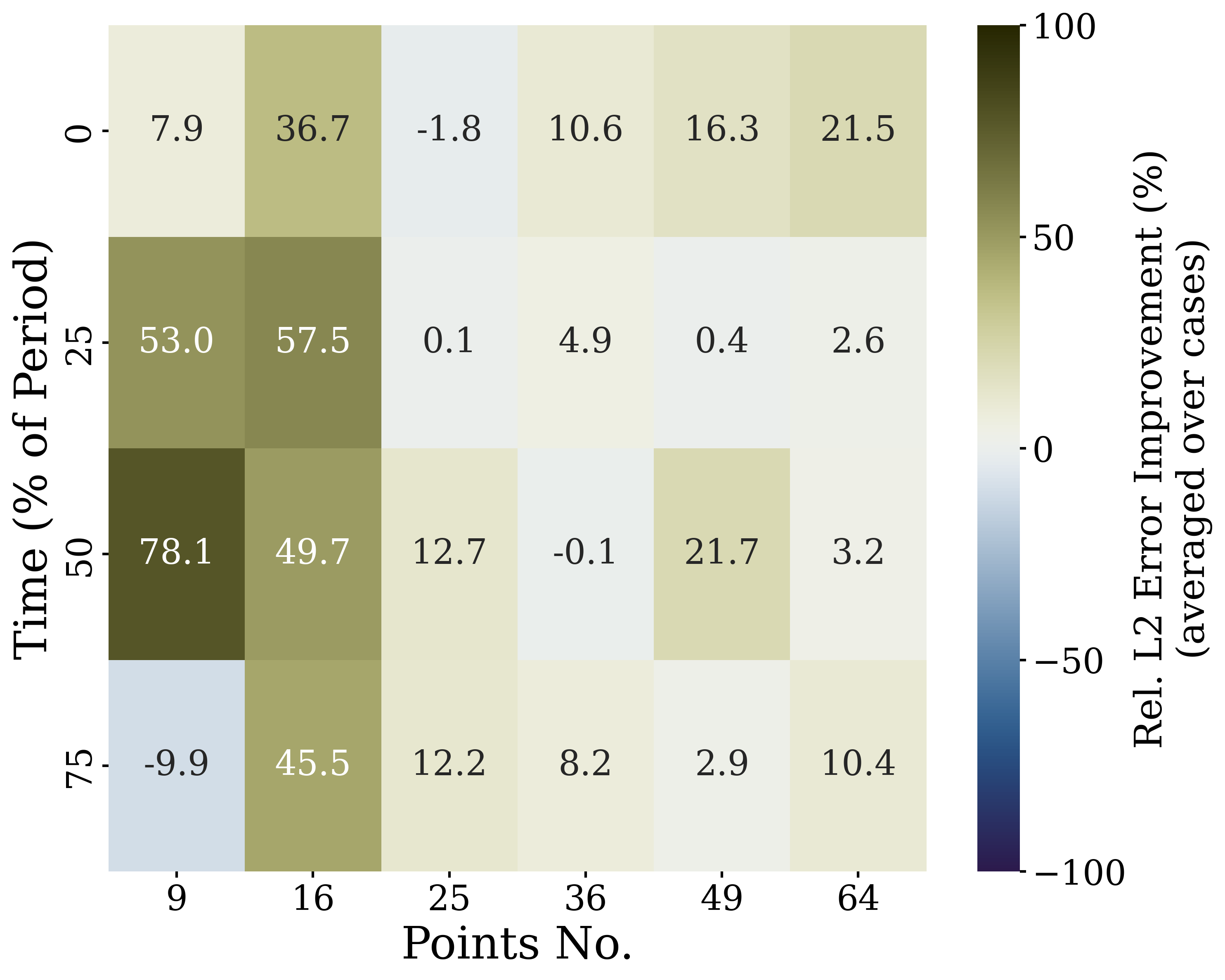}
        \subcaption{L2 error improvement for $\mathbf{v}_\text{y}$ component}
        \label{fig:vy_improvement_RUNavg}
    \end{subfigure}
    
    \caption{{\small Comparison of case-averaged L2 error between structured and optimized sensor placement strategies. Left column shows results for the $\mathbf{v}_\text{x}$ velocity component, right column shows results for the $\mathbf{v}_\text{y}$ velocity component. Top row: structured placement, middle row: optimized placement, bottom row: improvement achieved by optimization. All values represent averages across multiple test cases.}}
    \label{fig:RunAvgHeatmap}
\end{figure*}

\begin{figure*}
    \centering
    \begin{subfigure}[t]{0.46\textwidth}
        \centering
        \includegraphics[width=\linewidth,trim=0 0 0 0,clip]{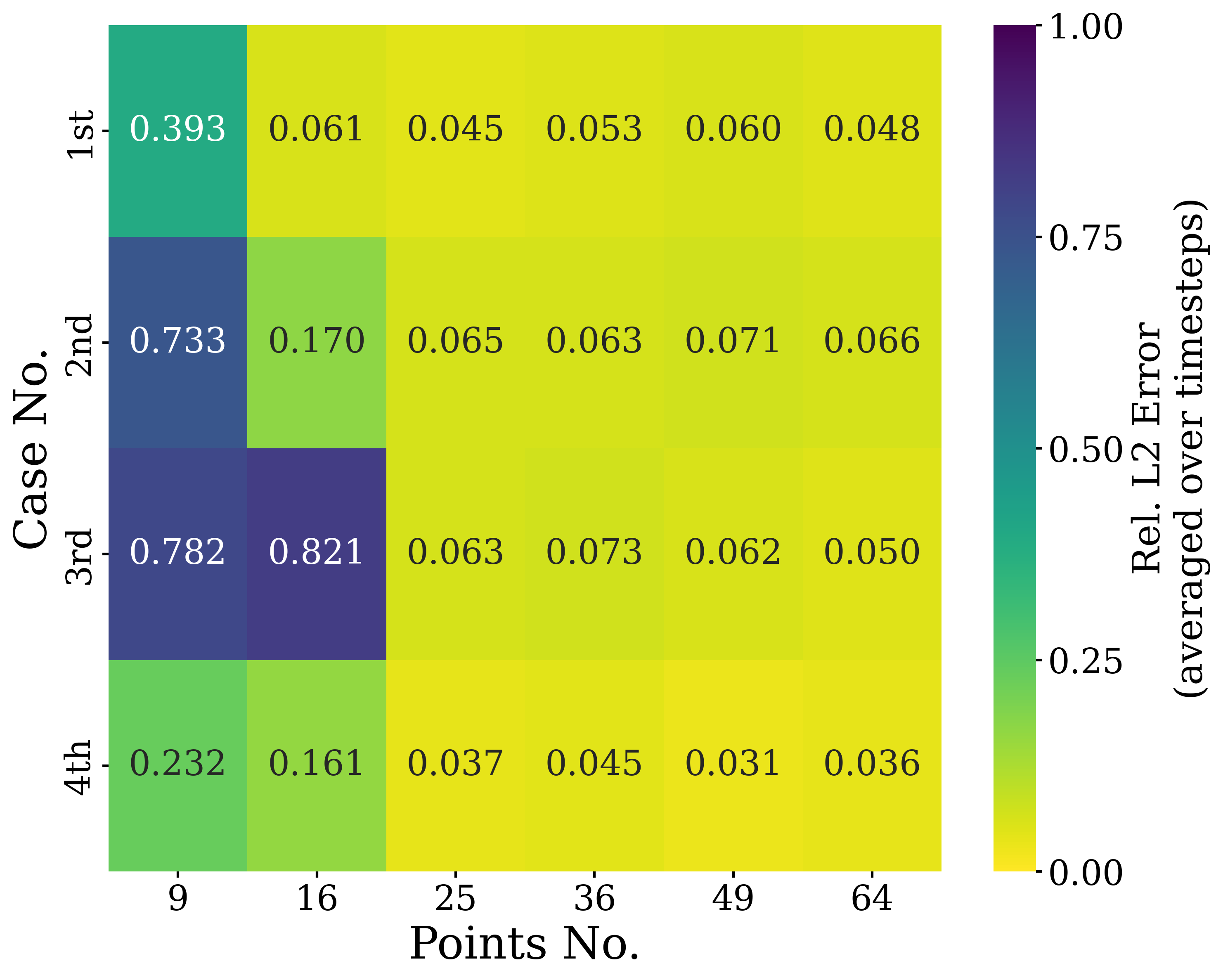}
        \subcaption{$\mathbf{v}_\text{x}$ L2 error with structured sensor placement}
        \label{fig:vx_structured_ntAvg}
    \end{subfigure}
    \hfill
    \begin{subfigure}[t]{0.46\textwidth}
        \centering
        \includegraphics[width=\linewidth,trim=0 0 0 0,clip]{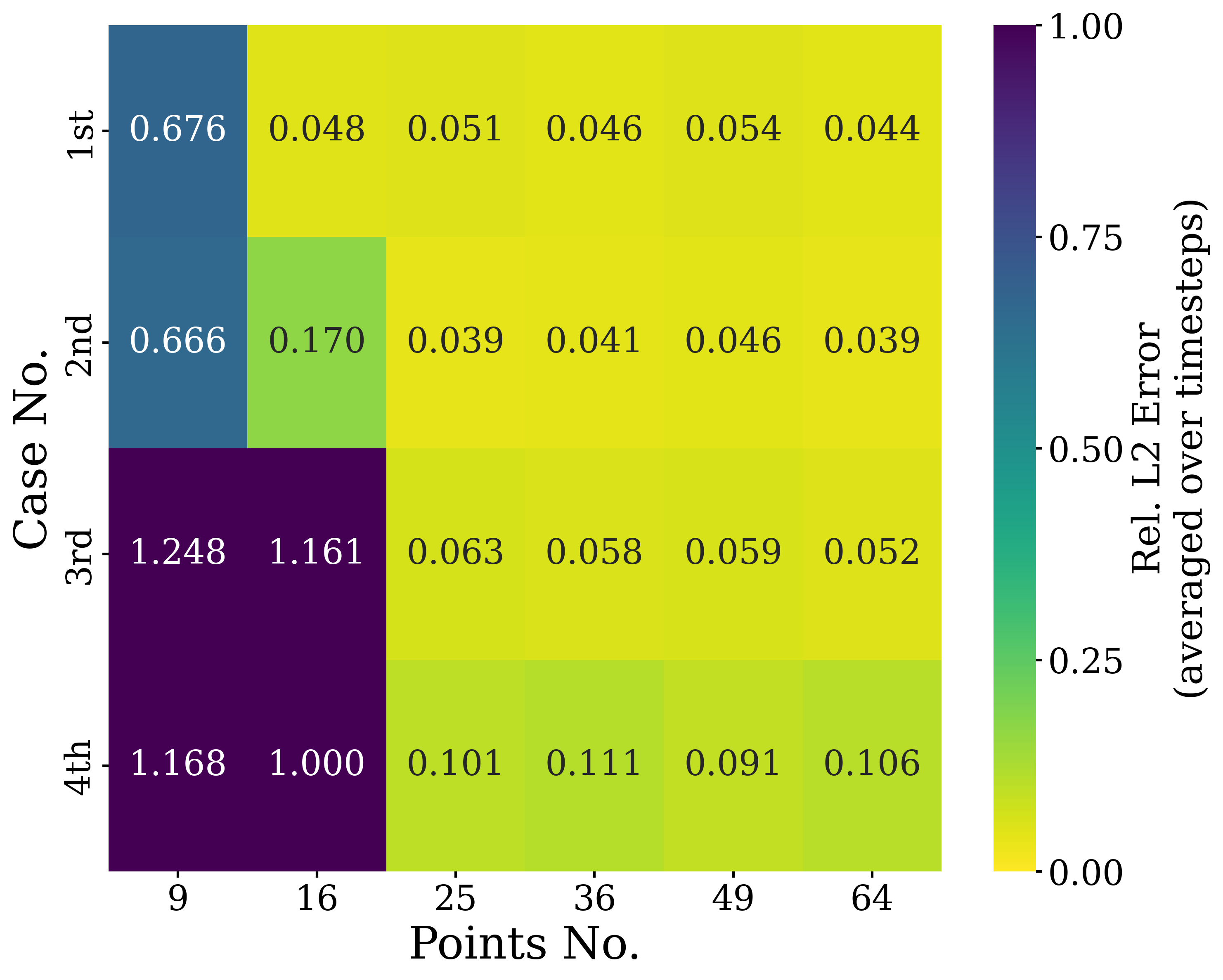}
        \subcaption{$\mathbf{v}_\text{y}$ L2 error with structured sensor placement}
        \label{fig:vy_structured_ntAvg}
    \end{subfigure}
    
    \vspace{0.5cm}
    
    \begin{subfigure}[t]{0.46\textwidth}
        \centering
        \includegraphics[width=\linewidth,trim=0 0 0 0,clip]{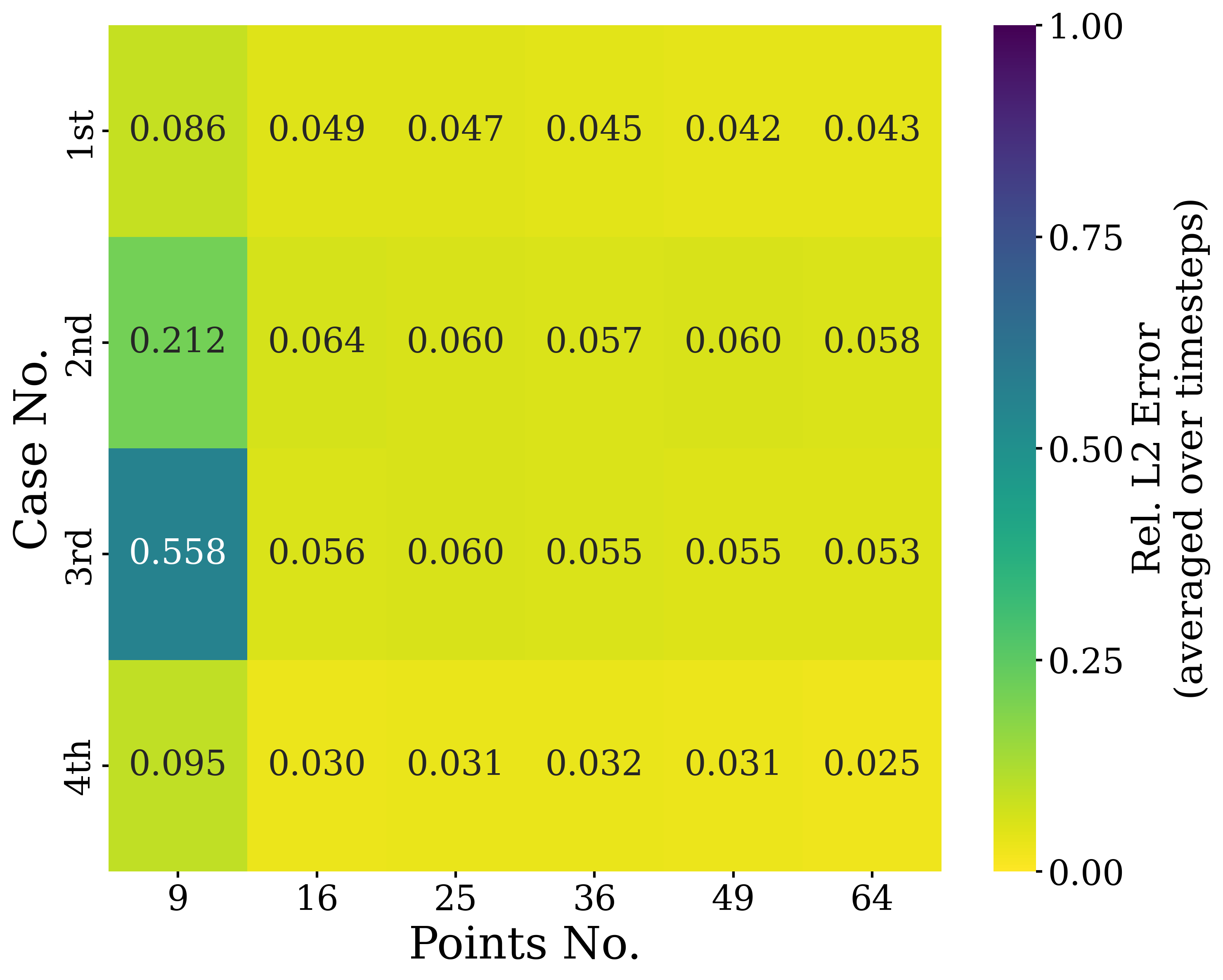}
        \subcaption{$\mathbf{v}_\text{x}$ L2 error with optimized sensor placement}
        \label{fig:vx_optimized_ntAvg}
    \end{subfigure}
    \hfill
    \begin{subfigure}[t]{0.46\textwidth}
        \centering
        \includegraphics[width=\linewidth,trim=0 0 0 0,clip]{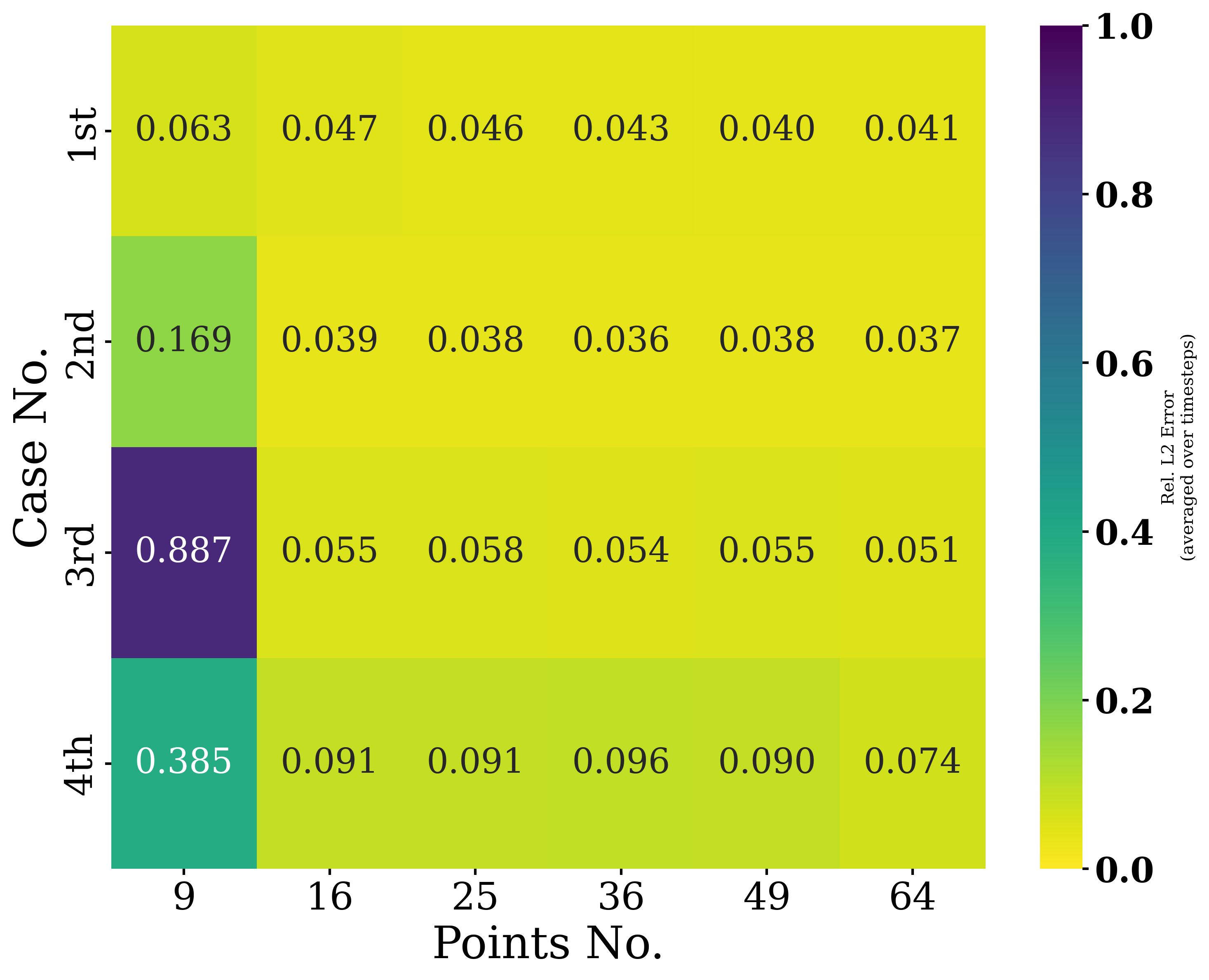}
        \subcaption{$\mathbf{v}_\text{y}$ L2 error with optimized sensor placement}
        \label{fig:vy_optimized_ntAvg}
    \end{subfigure}
    
    \vspace{0.5cm}
    
    \begin{subfigure}[t]{0.46\textwidth}
        \centering
        \includegraphics[width=\linewidth,trim=0 0 0 0,clip]{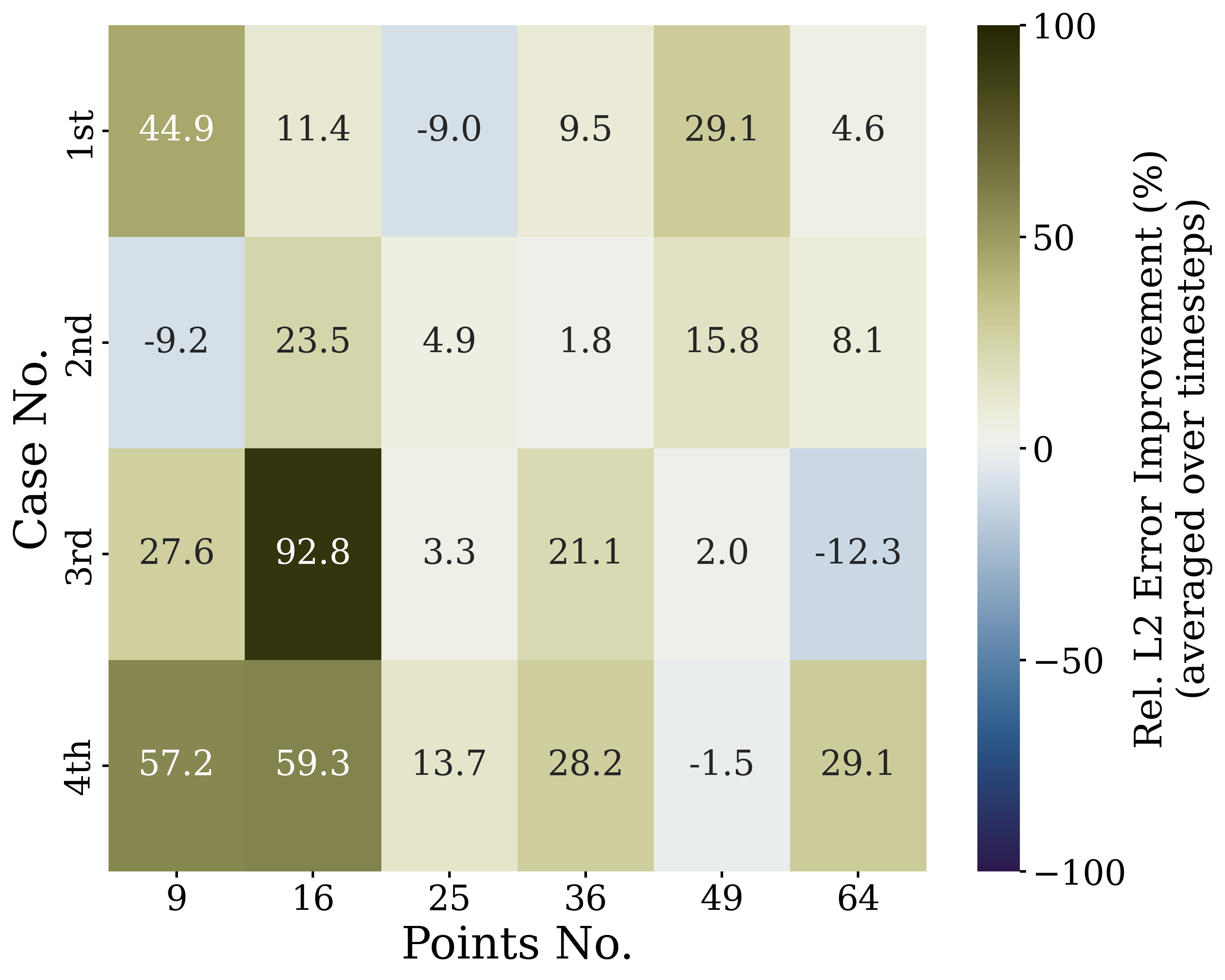}
        \subcaption{L2 error improvement for $\mathbf{v}_\text{x}$ component}
        \label{fig:vx_improvement_ntAvg}
    \end{subfigure}
    \hfill
    \begin{subfigure}[t]{0.46\textwidth}
        \centering
        \includegraphics[width=\linewidth,trim=0 0 0 0,clip]{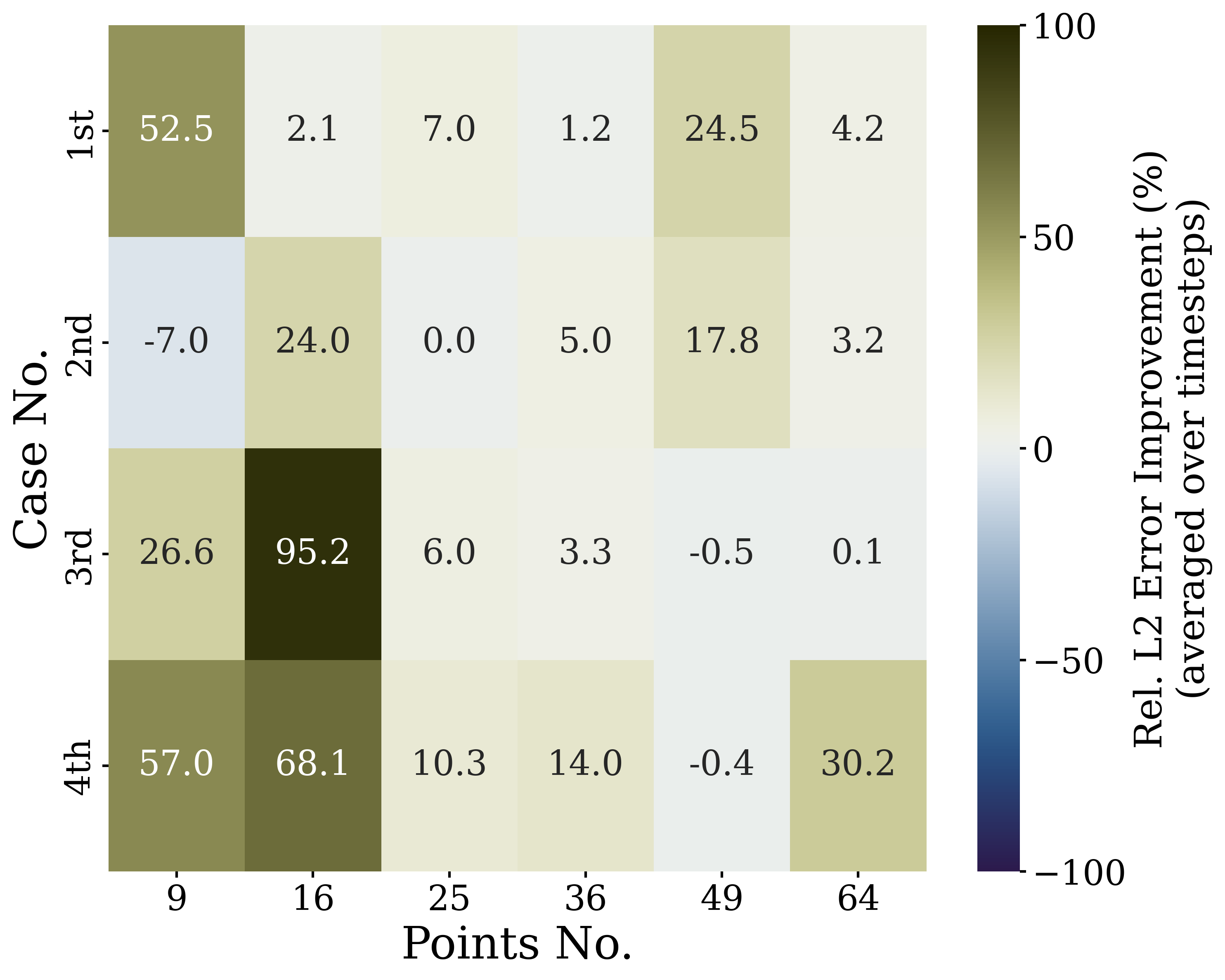}
        \subcaption{L2 error improvement for $\mathbf{v}_\text{y}$ component}
        \label{fig:vy_improvement_ntAvg}
    \end{subfigure}
    
    \caption{{\small Comparison of timestep-averaged L2 error between structured and optimized sensor placement strategies. Left column shows results for the $\mathbf{v}_\text{x}$ velocity component, right column shows results for the $\mathbf{v}_\text{y}$ velocity component. Top row: structured placement, middle row: optimized placement, bottom row: improvement achieved by optimization. All values represent averages across multiple timesteps.}}
    \label{fig:ntAvgHeatmap}
\end{figure*}

To improve reconstruction performance, we applied a greedy MI-based optimization procedure to identify the most informative sensor locations. Figure~\ref{fig:mutualinfo_points} shows the outcome of this process, where 64 sensor points are distributed throughout the domain in order of informativeness. Each point is annotated with its selection index during the greedy process and color-coded into eight groups for visual clarity. The earliest selected points (e.g., Points 1–4 in red) cluster around regions characterized by strong unsteady dynamics, such as vortex cores and wake interaction zones. These locations are known to carry high information content. As the selection progresses, later points (e.g., Points 50–64 in light pink) tend to fill in less informative or more redundant regions of the flow. This progression reflects the diminishing marginal gain in information as the sensor network becomes denser.

To evaluate the effectiveness of our sensor position optimization strategy, we compare the reconstruction performance between structured and optimized sensor placements. This comparison forms the basis of the following discussion.

To begin the performance assessment, Fig.~\ref{fig:barimprovement} presents a comprehensive comparison of the total average L2 errors in the x- and y-velocity components, aggregated across all test cases, timesteps, and spatial points. This global averaging provides a high-level summary of reconstruction accuracy and serves as a natural starting point for evaluation. The blue and green bars represent the performance of structured and optimized sensor placements, respectively, across varying numbers of observation points (ranging from 9 to 64).  The black error bars denote the standard deviation, reflecting variability across test scenarios. A consistent trend emerges in both subplots: optimized sensor placements outperform structured ones across all sensor counts. The most pronounced differences are observed at lower sensor counts, where the optimized strategy yields significantly lower errors. The percentage improvement of optimized over structured placement is annotated above each green bar, providing a direct quantitative comparison. 

The impact of optimized sensor placement is most striking at lower observation point counts. For instance, at 16 observation points, the optimized placement yields dramatic improvements of 83\% and 90\% for $\mathbf{v}_\text{x}$ and $\mathbf{v}_\text{y}$, respectively. This peak in improvement underscores the value of strategic sensor positioning when the number of measurements is limited. Similarly, with only 9 observation points, substantial improvements of 55\% and 60\% are observed, again highlighting the critical importance of sensor position optimization in data-scarce regimes.

As the number of observation points increases, the relative advantage of optimized placement diminishes. Beyond 25 sensors, the performance gap between structured and optimized placements narrows significantly. This is aligned with previous findings suggesting that reconstruction accuracy saturates with a sufficient number of observation points, regardless of their exact placement. In such regimes, the inherent information content captured by the sensors becomes adequate for the DDPM network to perform robustly, making optimization less impactful.

To further discuss the reconstruction performance and move beyond a fully aggregated view, we next examine the influence of test case variability and timestep on the L2 error distribution. The goal of this analysis is to assess how the reconstruction accuracy varies not only with the number of sensors but also as a function of test cases and timesteps. Toward this end, we compute two separate sets of spatially averaged L2 errors: one averaged across timesteps (to highlight case-wise effects) and one averaged across test cases (to reveal temporal trends).

Figs.~\ref{fig:RunAvgHeatmap}\subref*{fig:vx_structured_RUNavg}--\subref*{fig:vy_optimized_RUNavg} illustrate the heatmaps for the average L2 error computed by averaging across different test cases. The comparison between structured and optimized sensor placements clearly reveals the dependency of reconstruction accuracy on the number of observation points. As anticipated, fewer observation points (e.g., 9 and 16) correspond to significantly higher reconstruction errors, reflecting insufficient information for the DDPM network. Conversely, increasing the number of observation points substantially enhances the reconstruction accuracy. Specifically, from approximately 25 observation points onwards, L2 errors stabilize around relatively low values, approximately 0.05. The maximum errors for structured placements at this level were 0.054 and 0.068 for $\mathbf{v}_\text{x}$ and $\mathbf{v}_\text{y}$, respectively. In contrast, optimized sensor placement consistently delivered lower errors, reaching minima of 0.039 for $\mathbf{v}_\text{x}$ and 0.048 for $\mathbf{v}_\text{y}$.

The highest errors occurred for the minimal observation scenarios (9 points). Here, the structured sensor placements exhibited maximum errors of 0.816 ($\mathbf{v}_\text{x}$) and 1.372 ($\mathbf{v}_\text{y}$), particularly prominent at timestep 75, which might correspond to a pronounced phase of vortex shedding. Optimized placements significantly reduced these maximum errors to 0.362 and 0.577 for $\mathbf{v}_\text{x}$ and $\mathbf{v}_\text{y}$, respectively.

Figs.~\ref{fig:RunAvgHeatmap}\subref*{fig:vx_improvement_RUNavg}--\subref*{fig:vy_improvement_RUNavg}  quantify the percentage improvement in reconstruction accuracy due to optimized placement. This improvement is calculated using the following relation:
\begin{align}
&\text{Improvement (\%)} =\nonumber \\ 
&\frac{\text{L2 error (structured)} - \text{L2 error (optimized)}}{\text{L2 error (structured)}} \times 100
\label{eq:improL2}
\end{align}

The heatmaps clearly indicate substantial improvements, especially with fewer observation points. The highest improvements were observed for the 9-point scenario of the third test case, reaching up to 69\% for $\mathbf{v}_\text{x}$ and 78\% for $\mathbf{v}_\text{y}$. Notably, however, negative improvements are occasionally recorded, depicted by sharp shifts into the blue spectrum in the heatmap. This phenomenon arises because the ROM-informed mutual information sensor placement method identifies optimal fixed sensor positions based on the entirety of the dataset, inherently averaging over different timesteps. Consequently, in unsteady wake conditions, fixed sensor positions cannot guarantee optimal accuracy at every timestep. This limitation occasionally leads to reduced performance at specific timesteps, explaining the presence of negative improvements in certain scenarios.

Fig. \ref{fig:ntAvgHeatmap} presents a complementary evaluation, showing L2 errors averaged over timesteps within each test case. Generally, errors in this timestep-averaged analysis are slightly lower compared to case-averaged results (Fig. \ref{fig:RunAvgHeatmap}). Similar to previous observations, structured sensor placements again demonstrate pronounced errors at lower numbers of observation points (9 and 16). The optimized sensor placements, however, reduce these errors significantly. Remarkably, optimized placements exhibit notably lower errors at 16 observation points, further emphasizing the advantage of optimization.

\begin{figure*}
    \centering
    \includegraphics[width=1\linewidth]{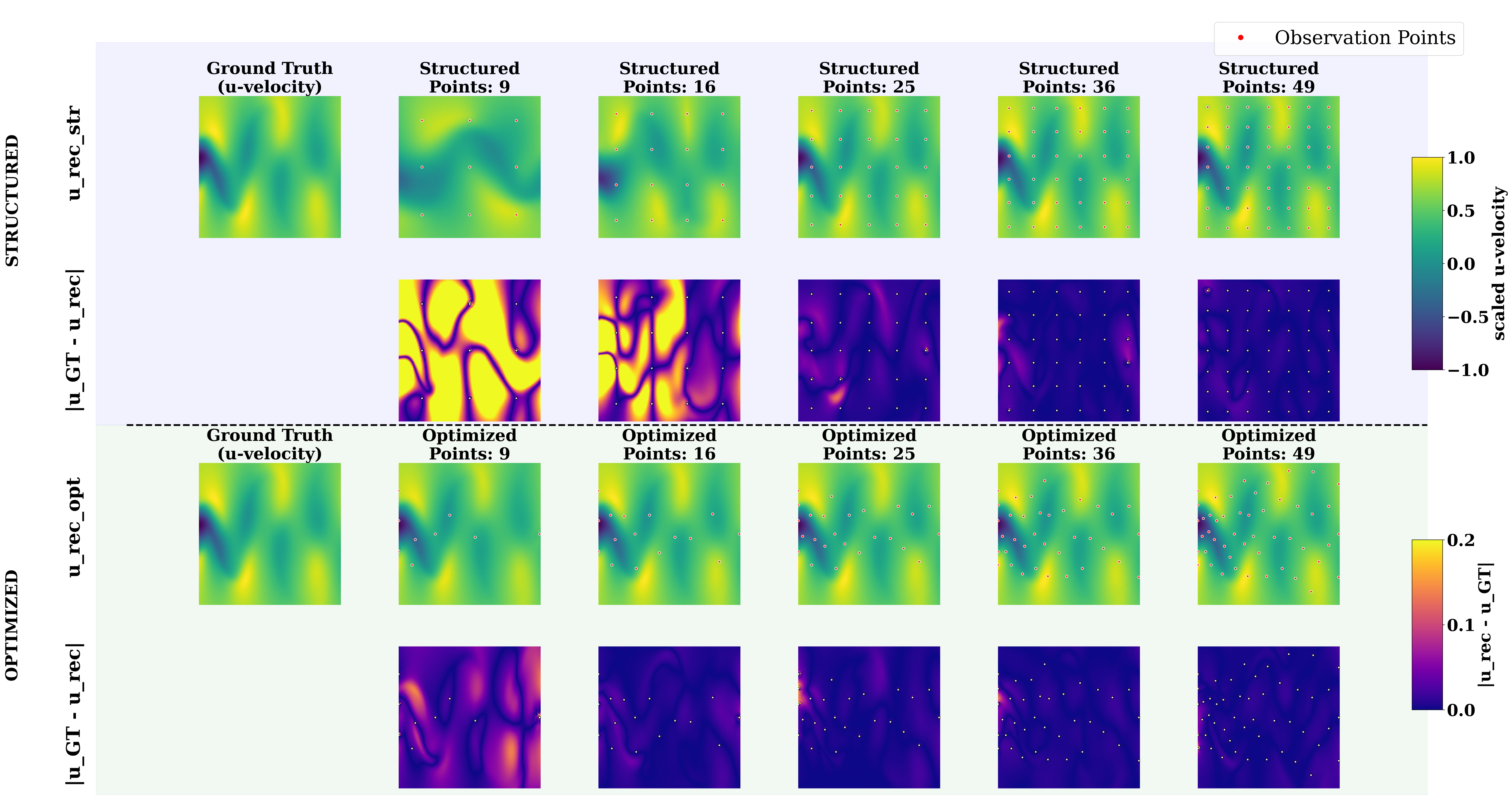}
    \caption{Visual comparison of ground truth, reconstructed fields, and corresponding error maps for $\mathbf{v}_\text{x}$ across different sensor counts. The top half shows results for structured sensor placements, while the bottom half corresponds to optimized placements. Red dots indicate sensor locations.}
    \label{fig:ContourComparisonu}
\end{figure*}

\begin{figure*}
    \centering
    \includegraphics[width=1\linewidth]{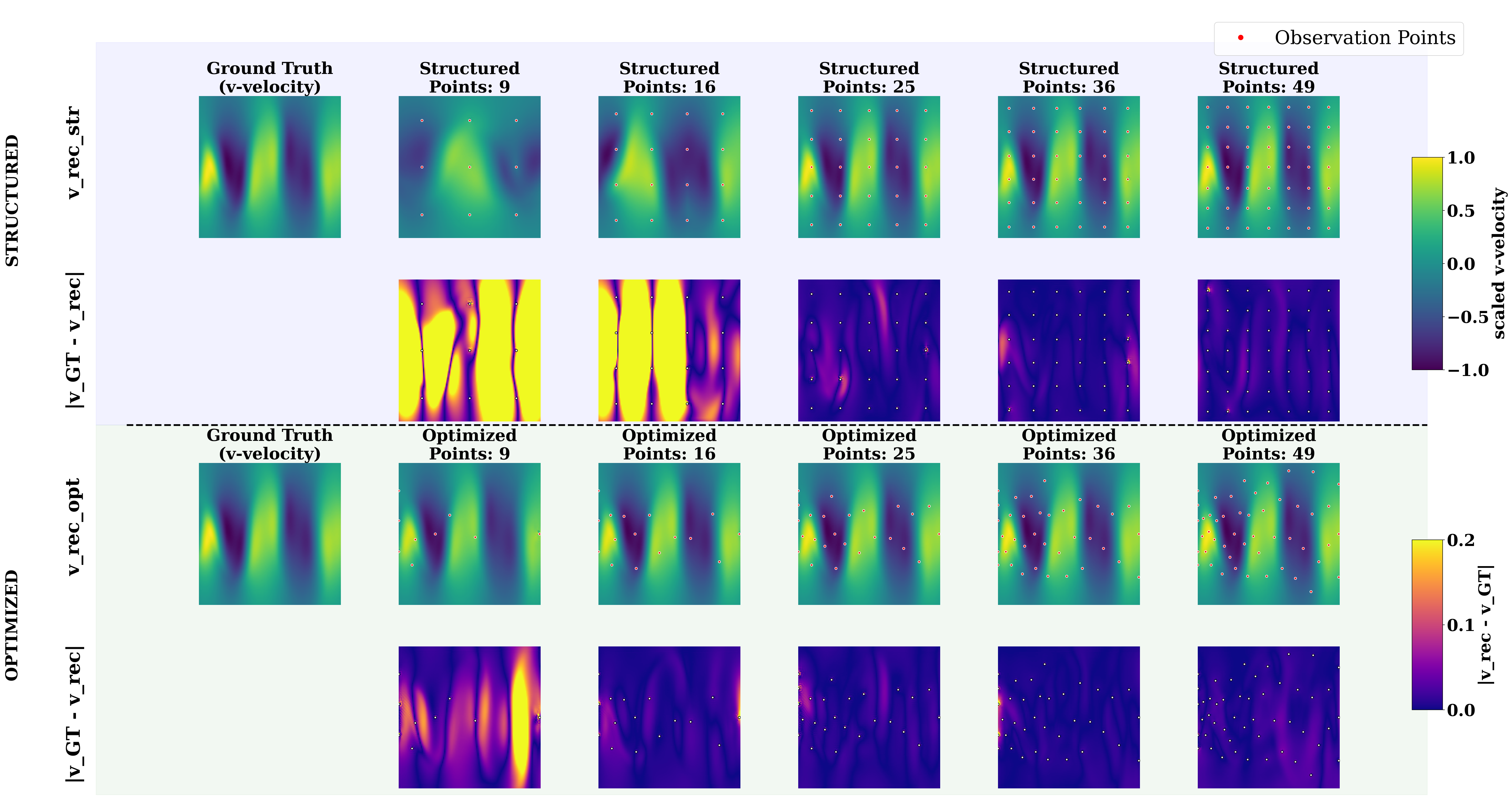}
    \caption{Visual comparison of ground truth, reconstructed fields, and corresponding error maps for $\mathbf{v}_\text{y}$ across different sensor counts. The top half shows results for structured sensor placements, while the bottom half corresponds to optimized placements. Red dots indicate sensor locations.}
    \label{fig:ContourComparisonv}
\end{figure*}

The maximum errors for structured sensor placements occur in the third test case, reaching 0.842 for $\mathbf{v}_\text{x}$ and 1.248 for $\mathbf{v}_\text{y}$. The optimized placements, however, show reduced errors of 0.558 and 0.887 for $\mathbf{v}_\text{x}$ and $\mathbf{v}_\text{y}$, respectively, translating to an approximately 27\% improvement as illustrated in Figs.~\ref{fig:ntAvgHeatmap}\subref*{fig:vx_improvement_ntAvg}--\subref*{fig:vy_improvement_ntAvg}. The significant improvement observed for this scenario underscores the efficacy of optimized sensor placement strategies, particularly under conditions of higher complexity and dynamic variability.

Finally, optimized sensor placements clearly demonstrate their effectiveness, achieving low errors consistently beyond 16 observation points. 
The guided DDPM approach, complemented by an optimized sensor placement strategy, significantly enhances the accuracy of reconstructing unsteady wake flows, especially in scenarios characterized by fewer observation points. This analysis emphasizes the value of optimization methodologies in practical applications involving fluid dynamics simulations and sensor deployments.

Figs. \ref{fig:ContourComparisonu} \& \ref{fig:ContourComparisonv} present the contour plot comparison of the reconstruction performance for each velocity component using structured and optimized sensor placements. The top half of the figure corresponds to structured placement, while the bottom half shows results for the optimized configuration. Each sub-panel includes the ground truth, reconstructed velocity field, and the associated L2 error map across different numbers of observation points (ranging from 9 to 49), to assess the reconstruction fidelity evolution with sensor density. It can be seen that the reconstruction accuracy improves consistently as the number of observation points increases. However, the advantage of optimized sensor placement is particularly clear at lower sensor counts (9 and 16), where reconstructions more closely resemble the ground truth and exhibit notably reduced spatial errors. In contrast, the structured approach yields visibly higher error concentrations, especially in the regions far from the observation. The L2 error contours corroborate this, highlighting substantial reductions in local reconstruction errors when using optimized placement. This advantage diminishes as sensor counts increase, with both strategies converging in performance beyond approximately 25 sensors, showing a saturation in error reduction. Additionally, the optimized sensor locations tend to cluster around high-information regions, whereas structured sensors are uniformly spaced, further explaining the efficiency gains in the optimized cases. 

\section{\label{sec5}Conclusion}
This work presented a dual-guided data-driven framework for reconstructing unsteady incompressible flows, where guidance is applied at two levels: first, in optimizing sensor placement, and then in physics-informed generative modeling via guided diffusion. The core contribution lies in the efficient identification of optimal sensor locations using mutual information theory applied to a reduced-order solution manifold. This method efficiently identifies sensor locations that maximize information gain, significantly reducing computational cost compared to brute-force or full-order approaches.

The optimized sensor configuration was then integrated into a guided DDPM framework, where the generative process was conditioned both on sparse observations and partial PDE knowledge. Our results clearly showed that in regimes with very limited observations (e.g., 9 or 16 sensors), structured sensor placements failed to provide reliable reconstructions, with high L2 errors exceeding 0.8 in some cases. In contrast, the optimized sensor placement yielded accurate reconstructions with L2 errors around 0.05, demonstrating robustness even under unsteady data conditions. As the number of sensors increased beyond 25, the difference between structured and optimized placements diminished, indicating a saturation regime where observation density compensates for suboptimal placement.

Looking ahead, several directions can further extend the capabilities of the proposed framework. One natural extension is to apply the methodology to more complex fluid scenarios, such as turbulent or three-dimensional flows, which pose greater challenges in both sensor placement optimization and generative reconstruction. Additionally, exploring alternative generative modeling paradigms, such as latent diffusion \cite{zhou2024text2pde} or flow-matching models \cite{baldan2025flow}, could offer improvements in training speed and sampling efficiency.

\section*{ACKNOWLEDGMENTS}
This work was supported by a research grant (VIL57365) from VILLUM FONDEN.
\section*{AUTHOR DECLARATIONS}
\subsection*{Conflict of Interest}
The authors have no conflicts to disclose.

\subsection*{Author Contributions}
SS: Conceptualization; Data curation; Formal analysis; Investigation; Methodology; Software; Visualization; Writing – original draft. HK: Formal analysis; Project administration; Resources; Supervision; Writing – review \& editing. AI: Conceptualization; Formal analysis; Methodology; Supervision; Writing – review \& editing. MA: Conceptualization; Formal analysis; Funding acquisition; Methodology; Project administration; Resources; Supervision; Writing – review \& editing.

\section*{DATA AVAILABILITY}
The data that support the findings of this study are available from the corresponding author upon reasonable request.

\appendix

\section{Full-Order Model Computational Settings}
The Full-Order Model (FOM) simulations were performed on a workstation equipped with two Intel(R) Xeon(R) E5-2680 v4 2.40 GHz processors and 64 GB of RAM. Each FOM run was executed using 2 CPU cores and required approximately 8 hours to reach a fully periodic field and complete the necessary sampling.

\section{Deep Learning Computational Settings}
This work builds directly upon the diffusion framework presented in Ref.\cite{huang2024diffusionpde}, specifically extending the ideas proposed for PDE-constrained generative modeling using denoising diffusion. For the core architecture, we adopt the publicly available implementation from NVIDIA’s EDM repository \cite{Karras2022edm}, which explores the design space of diffusion-based generative models and offers a modular and robust training pipeline.

Our experiments use the \texttt{ddpmpp} configuration (Denoising Diffusion Probabilistic Models++) with EDM-style preconditioning. The generator is a modified U-Net architecture, referred to as \texttt{SongUNet}, with positional embedding and a standard encoder-decoder structure. Key architecture settings include:

\begin{itemize}
    \item \textbf{Channels}: Initial base of 128 channels, scaled by a multiplier \texttt{[2, 2, 2]} across layers.
    \item \textbf{Dropout}: 0.13 dropout rate applied during training.
    \item \textbf{Augmentations}: Training incorporates moderate data augmentations with probability 0.12, including flipping, translation, and rotation.
\end{itemize}

The deep learning model was trained using two NVIDIA A6000 GPUs. Training the diffusion model took approximately two weeks. During inference, each sampling process required around 5$\sim$10 minutes.

\bibliography{ref}

\end{document}